\documentclass{article}

\usepackage{arxiv}

\usepackage[utf8]{inputenc} % allow utf-8 input
\usepackage[T1]{fontenc}    % use 8-bit T1 fonts
\usepackage{hyperref}       % hyperlinks
\usepackage{url}            % simple URL typesetting
\usepackage{booktabs}       % professional-quality tables
\usepackage{amsfonts}       % blackboard math symbols
\usepackage{nicefrac}       % compact symbols for 1/2, etc.
\usepackage{microtype}      % microtypography
\usepackage{lipsum}
\usepackage{graphicx}
\usepackage[fleqn]{amsmath}
\usepackage{algorithm}
\usepackage{algpseudocode}

\title{Preserving spreading dynamics and information flow in complex network reduction}

\author{
 Dan Chen \\
   School of Mathematics and Statistics \\
   Wuhan Textile University \\
   Wuhan 430200, China \\
  \texttt{danchen@wtu.edu.cn} \\
   \And
 Housheng Su \\
   School of Artificial Intelligence and Automation \\
   Huazhong University of Science and Technology \\
   Wuhan 430074, China \\
  \texttt{houshengsu@gmail.com} \\
  \And
 Yong Wang \\
   School of Artificial Intelligence and Automation \\
   Huazhong University of Science and Technology \\
   Wuhan 430074, China \\
  \texttt{yongwang@hust.edu.cn} \\
  \And
 Jie Liu \\
   School of Mathematics and Statistics \\
   Wuhan Textile University \\
   Wuhan 430200, China \\
  \texttt{liujie@wtu.edu.cn} \\
}

\begin{document}
\maketitle
\begin{abstract}
Effectively preserving both the structural and dynamical properties during the reduction of complex networks remains a significant research topic. Existing network reduction methods based on renormalization group or sampling often face challenges such as high computational complexity and the loss of critical dynamic attributes. This paper proposes an efficient network reduction framework based on subgraph extraction, which accurately preserves epidemic spreading dynamics and information flow through a coordinated optimization strategy of node removal and edge pruning. Specifically, a node removal algorithm driven by enhanced degree centrality is introduced to preferentially remove low-centrality nodes, thereby constructing a smaller-scale subnetwork. Subsequently, an edge pruning algorithm is designed to regulate the edge density of the subnetwork, ensuring that its average degree remains approximately consistent with that of the original network. Experimental results on Erd\"os-R\'enyi random graphs, Barab\'asi-Albert scale-free networks, and real-world social contact networks from various domains demonstrate that this proposed method can reduce the size of networks with heterogeneous structures by more than 85\%, while preserving their epidemic dynamics and information flow. More importantly, our method almost always achieves the highest accuracy compared to state-of-the-art techniques. These findings provide valuable insights for predicting the dynamical behavior of large-scale real-world networks, and also reveal that a large number of nodes and edges in real-world networks play redundant roles in information transmission.
\end{abstract}

% keywords can be removed

%%%%%%%%%%%%%%%%%%%%%%%%%%%%%%%%%%%%%%%%%%%%%%%%%%%%%%
%%% The main text.
%%%%%%%%%%%%%%%%%%%%%%%%%%%%%%%%%%%%%%%%%%%%%%%%%%%%%%%
\section{Introduction}
Networks are essential tools for understanding large-scale complex systems and are utilized across various disciplines, including physics, sociology, biology, information science, and computer science~\cite{Barabasi2016}. However, real-world network systems often exhibit intricate interactions, diverse structures, and large scales~\cite{Newman2018,Martin2019}, which can make analyzing dynamic processes—such as disease and information propagation~\cite{Wang2019,Chen2023Identification}, synchronization~\cite{Arenas2008,Koronovskii2017}, reaction-diffusion~\cite{Colizza2007,Gomez2013}, and controllability~\cite{Liu2011,Su2023}—quite challenging. One promising approach to tackle these challenges leverages concepts and techniques from renormalization group (RG) theory~\cite{Song2005,Serrano2008,Garcia2018,Koch2018,Chen2021,Chen2023a,Garuccio2023,Klemm2023,Villegas2023,loures2023,Chen2023Scaling,Kolk2024,Li2024predicting,Poggialini2025,Zhang2025,Nurisso2025,Gabrielli2025} and sampling~\cite{Leskovec2006,Lee2006,Yoon2007,Hubler2008,Li2019,Zhao2021,Zhu2021DRGraph,ChenW2022,Jiao2023,Zhang2023,Chen2023b,Chen2024,Zhou2025}. The RG approach systematically reduces the scale of networks by successively coarse-graining nodes and links. Representative methods include box-covering renormalization~\cite{Song2005}, geometric renormalization~\cite{Garcia2018}, and Laplacian renormalization~\cite{Villegas2023}. However, these methods often rely on domain-specific assumptions or incur high computational complexity, which limits their scalability for large network systems.

Compared with renormalization methods, subgraph extraction strategies that focus on sampling techniques~\cite{Leskovec2006,Lee2006,Yoon2007,Hubler2008,Li2019,Zhao2021,Zhu2021DRGraph,ChenW2022,Jiao2023,Zhang2023} bypass the limitations of many preset conditions. Importantly, these approaches typically have low time complexity, which is crucial for handling large-scale networks efficiently. However, while the current mainstream sampling techniques pursue efficiency advantages, they also present several challenges. For example, node or edge-sampling-based methods~\cite{Lee2006} often struggle to accurately preserve the core characteristics of the original network in the resulting subgraphs. Furthermore, the inherent probabilistic nature of sampling methods means that identical sampling schemes can yield significantly different results across different implementations.

In this context, this paper focuses on simplifying the topological structure of complex networks through node removal and edge pruning strategies. Compared to the coarse-graining method based on renormalization, our approach has significantly lower computational complexity. Specifically, we present the node's enhanced degree centrality (DC$_{+}$). The order in which nodes are removed is inversely proportional to their DC$_{+}$ values, meaning that nodes with lower DC$_{+}$ values are preferentially removed from the original network. By setting a preset node removal ratio, we can flexibly reduce the size of the subnetwork to any manageable level. Another purpose of this setting is to facilitate the comparative analysis among different reduction methods. We investigate the epidemic spreading dynamics of synthetic and real-world networks under the proposed node removal method. The results indicate that the node removal process is often accompanied by an increase in average degree, which enables the subgraphs to exhibit stronger spreading abilities than the original network. To mitigate the effects of the high average degree, we introduce an edge pruning algorithm designed to adjust the edge density of the subgraph to a level comparable to that of the original network. Simulation experiments conducted on two types of synthetic networks and some real networks demonstrate that, for most networks with heterogeneous structures, the subgraphs obtained through the combined effects of node removal and edge pruning can accurately reproduce the epidemic spreading dynamics process in the original network. Additionally, we further study the evolution behavior of network information dynamics based on Laplacian diffusion during the subgraph extraction process. Interestingly, our method also retains the information flow of the original network, similar to the recent coarse-graining method~\cite{Zhang2025}, suggesting that the nodes and edges discarded during the subgraph extraction process play a redundant role in information transmission.

The main contributions of the paper can be summarized as follows:

	$\bullet$ This paper designs a method for extracting network topology subgraphs that integrates node removal and edge pruning. This approach effectively reduces the topological scale of the network and is significantly faster—by several orders of magnitude—than existing coarse-graining methods in terms of computational complexity.
	
	$\bullet$ Experiments conducted on real networks across various fields show that the proposed method accurately preserves the spreading dynamics and information flow of the original network while compressing the scale of the original network by more than 85\%. This finding offers new practical guidance for predicting complex dynamic processes and their critical parameters on large-scale real networks.
	
	$\bullet$ We conducted extensive evaluations of our proposed method alongside other mainstream sampling-based subgraph extraction techniques, confirming the superiority of our method in preserving spreading dynamics and information flow.

The organization of this paper is structured as follows: Section~\ref{2} provides a brief overview of relevant network reduction works. In Section~\ref{3}, we introduce the node removal strategy based on DC$_{+}$ and propose an edge pruning algorithm, along with a brief analysis of the algorithm's computational complexity. Section~\ref{4} first outlines the basic information of the network datasets used in this study, then conducts an in-depth examination of the evolution of these networks' average degree and largest connected component during the node removal process. In particular, the spreading dynamics and the information flow of these networks during the reduction process are further studied. On this basis, the proposed method is compared with other mainstream reduction methods, followed by a brief elaboration on the potential application scenarios and research significance of network reduction. Finally, the main conclusion of the full paper is given in Section~\ref{5}.

\section{Related Works}\label{2}
Current methods for reducing complex networks can be divided into two main categories: iterative coarse-graining techniques based on renormalization group theory and sampling-based subgraph extraction methods. This section will review the relevant research from both perspectives.

\subsection{Network reduction algorithms based on RG}\label{2.1}
The RG theory provides a powerful framework for studying the self-similarity or scale invariance of complex networks at different resolutions. Below is a succinct overview of notable methods based on RG: Based on shortest path length-driven coarse-graining, a power-law relationship between the number of boxes required to cover a network and the box size was discovered, with a finite self-similarity exponent defined to elucidate network self-similar properties~\cite{Song2005}. Geometric renormalization~\cite{Garcia2018} (GR) embeds networks into a hidden metric space and aggregates adjacent nodes into supernodes via new mappings, revealing geometric scaling in real-world scale-free networks. This approach allows the networks to unfold into self-similar multi-layer shells that distinguish coexisting scales and their interactions. Additionally, they also developed high-fidelity network replicas and multi-scale navigation protocols. The Laplacian renormalization group~\cite{Villegas2023} (LRG) leverages diffusion dynamics and Kadanoff supernodes for momentum-space coarse-graining, analyzing scale invariance through network density matrices and entropy susceptibility to identify multi-scale structures and correlate spectral dimensions with entropy loss rates in both synthetic and real networks. Laplacian coarse-graining~\cite{loures2023} (LCG) reduces network topological scale, demonstrating cross-scale self-similarity in numerous artificial and real-world networks. DiskNet~\cite{Li2024predicting} integrates hyperbolic geometric renormalization with graph neural ordinary differential equations, identifying network skeletons via physics-informed embedding and mapping dynamics using degree-based super-resolution modules, outperforming traditional methods in multi-class network dynamic prediction~\cite{Li2024predicting}. In the context of LRG framework, scale-invariant networks were redefined by constant entropy loss rates ~\cite{Poggialini2025}, distinguishing scale-free from scale-invariant networks via constant entropy loss rates. Compiling a list of natural and artificial networks, they found significant scale invariance in the human connectome, revealing the correlation between spectral dimension and scale invariance. The network flow compression model~\cite{Zhang2025} fuses statistical physics and machine learning, employing graph neural networks to learn node grouping strategies while preserving information flow dynamics via partition function minimization. Recently, RG was extended to higher-order networks via cross-order Laplacian operators~\cite{Nurisso2025}, defining higher-order scale invariance and coarse-graining schemes that detected scale invariance at specific orders in pseudo-fractal and real higher-order networks, providing multi-order analysis tools for complex systems. More recently, a comprehensive review~\cite{Gabrielli2025} summarized major challenges, methods, and advances in network renormalization. The review covers geometric, Laplacian, and multi-scale approaches, highlighting limitations, advocating a first-principles universal framework, and discussing its multidisciplinary applications.

\subsection{Network reduction algorithms based on subgraph extraction}\label{2.2}
Graph sampling is a crucial technology for reducing complex networks and has seen significant advancements in recent years. Large-scale graph sampling methods have been investigated ~\cite{Leskovec2006}, comparing multiple algorithms and proposing evaluation criteria and scaling laws, and concluded that methods like random walks perform optimally. The impacts and biases of node, link, and snowball sampling on the topological properties of scale-free networks have also been examined~\cite{Lee2006}, revealing that each sampling method uniquely affects network topological features and explaining why these sampling processes can lead to estimation biases in various metrics. Ref~\cite{Yoon2007} explored the effects of random walk sampling on network statistical properties, revealing that scale-free networks with degree distribution exponent $\lambda \leq 3$ retain topological properties similar to the original networks, while significant biases emerge when $\lambda > 3$. The Metropolis algorithm~\cite{Hubler2008} (MHRW) method generates representative subgraphs that preserve the original topology through Markov chain Monte Carlo optimization. The Common Neighbor-Aware Random Walk~\cite{Li2019} (CNARW) method leverages edge weights and common neighbor information to accelerate convergence and enhance sampling efficiency. A minority-centric graph sampling~\cite{Zhao2021} (MCGS) method aimed at preserving minority structures. This approach balances the retention of minority and majority structures by identifying super-hubs and giant stars, while also optimizing sampling strategies. In~\cite{ChenW2022}, the impacts of random node sampling, breadth-first search, and hybrid methods on the topological properties of complex networks have been analyzed, demonstrating that sampling biases result from the combined effects of network heterogeneity and method-specific biases. A hierarchical sampling method for visualizing larger scale-free graphs was introduced~\cite{Jiao2023}, constructing a Core-Vertical-Graph-Periphery model to partition graphs into the core, vertical graph, and periphery components. The Hierarchical Structure Sampling (HSS) algorithm preserves core high-degree node connections, vertical graph joint degree distributions, and periphery low-degree node connection ratios. The effectiveness of HSS in sampling million-node graphs is validated through assessments of global statistical properties and local visual features. In parallel, another study~\cite{Zhang2023} proposed two novel top-level leadership sampling methods (TLS-e and TLS-i) that retain graph clustering structures by selecting top-layer nodes and expanding neighborhoods. Recently, we proposed a subgraph extraction method based on Edge-Reinforced Random Walk~\cite{Chen2024}, this approach yields high-fidelity smaller-scale subgraphs that retain the original network's structural properties through iterative edge-weight reinforcement sampling.

Despite recent advancements pushing complex network reduction research to new heights, critical bottlenecks remain to be addressed, such as reducing computational complexity and enhancing the fidelity of dynamical properties.

\section{Proposed Method}\label{3}
\begin{figure*}[htbp]
\begin{center}
\includegraphics[width=1.0\linewidth]{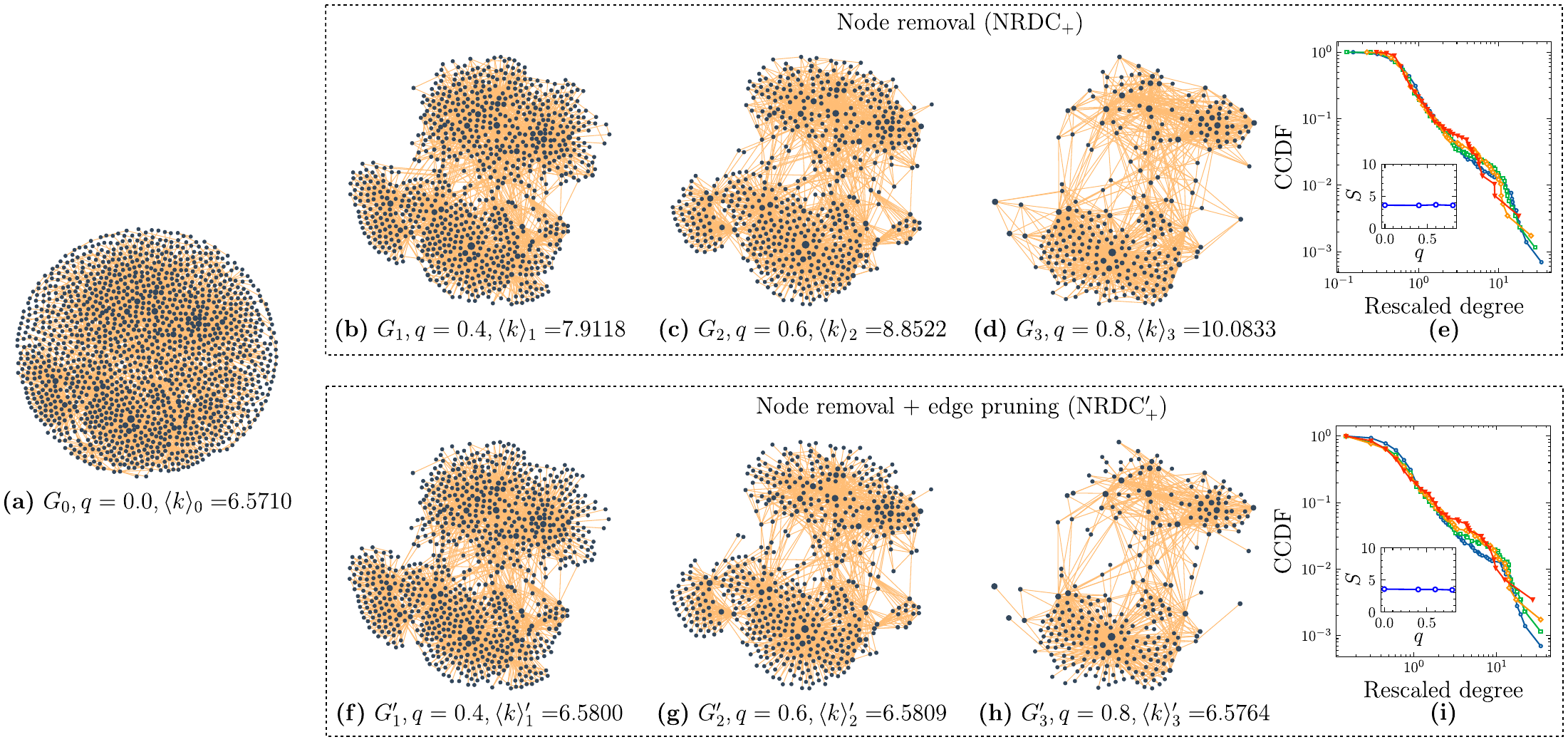}
\caption {Schematic illustration of network reduction via node removal and edge pruning. (a) The initial Metabolic network $G_0$. (b)-(d) Three subnetwork snapshots obtained from $G_0$ ($q$ = 0.4, 0.6, and 0.8), denoted as $G_1$, $G_2$, and $G_3$, respectively. (e) Shows the complementary cumulative degree distribution (CCDF) of $G_0$ and three subnetworks, the inset displays the information entropy $S$ corresponding to their degree distributions. (f)-(i) Present the results after edge pruning. The results demonstrate that the Metabolic network exhibits significant self-similarity in both CCDF and $S$ under the NRDC$_{+}$ and NRDC$_{+}^\prime$ methods.}
\label{Fig:1}
\end{center}
\end{figure*}

The primary task of this study is to effectively reduce the topological scale of complex networks through subgraph extraction, while maintaining the integrity of the network's spreading dynamics and information flow. Specifically, the proposed reduction algorithm consists of the following two main parts:

\textbf{(1) Node removal:} For the initial network $G_0$, we first calculate the enhanced degree DC$_{+}$ of all nodes and store these results in a sequence $\mathcal{S} = \{DC_{+}(1), DC_{+}(2), \dots, DC_{+}(N)\}$, where $N$ denotes the total number of nodes in $G_0$. The enhanced degree value of node $i$, denoted as DC$_{+}(i)$, is defined as the product of its degree value and the average neighbor degree~\cite{Chen2023Identification}. This sequence $\mathcal{S}$ is then sorted in ascending order to obtain a new sequence $\mathcal{S}^{\prime}$. Based on this sorted sequence, we select the nodes with the smallest DC$_{+}$ values in $\mathcal{S}^{\prime}$ and remove them from $G_0$ according to a preset removal ratio $q$. This process effectively reduces the scale of the network topology. Figures~\ref{Fig:1}(b)-(d) visually present three subgraph snapshots $G_1$, $G_2$, and $G_3$, which have been extracted from the Metabolic network shown in Figure~\ref{Fig:1}(a) at removal ratios of $q = 0.4$, $0.6$, and $0.8$, respectively. These subgraphs contain fewer nodes and edges than the original network. In the subsequent experiments, we named the node removal method based on DC$_{+}$ as NRDC$_{+}$. In Supplementary File, we demonstrate the differences between node degree centrality (DC) and DC$_{+}$ (see Figures~A1 and A2). It can be observed that DC$_{+}$ can better distinguish the differences among nodes, which is the motivation for adopting DC$_{+}$ to remove nodes in this paper.

\textbf{(2) Edge pruning:} In many heterogeneous networks, the subgraphs generated in step (1) often exhibit higher edge densities compared to the original network. Our subsequent experiments show that subgraphs with higher average degrees tend to have stronger spreading capabilities. To ensure that the subgraphs not only reflect a reduced topological scale but also preserve the spreading dynamic behavior of the original network, we introduce an edge pruning algorithm (see the pseudocode in Algorithm~\ref{alg:1}). This algorithm adjusts the edge density of the subgraphs to levels comparable to those of the original network, allowing us to evaluate whether the pruned subgraphs can effectively replicate the spreading dynamics of the original network. Taking the Metabolic network as an example, its average degree is $\langle k \rangle_0 = 6.5710$. The three corresponding subgraphs generated under node removal ratios of $q = 0.4$, $0.6$, and $0.8$ have average degrees of $\langle k \rangle_1 = 7.9118$, $\langle k \rangle_2 = 8.8522$, and $\langle k \rangle_3 = 10.0833$, respectively, as shown in Figures~\ref{Fig:1}(b)-(d). The result indicates that the average degree of the subgraphs increases with the removal ratio $q$. Subsequently, edge pruning is applied to these subgraphs, and the pruned versions are displayed in Figures~\ref{Fig:1}(f)-(h). After pruning, the average degrees are adjusted to $\langle k \rangle_1^{\prime} = 6.5800$, $\langle k \rangle_2^{\prime} = 6.5809$, and $\langle k \rangle_3^{\prime} = 6.5764$, which are very close to that of the original Metabolic network. In subsequent experiments, for cases where the density of the subgraph exceeds that of the original network, we will perform edge pruning to obtain a version with a similar edge density to the original network. We refer to the entire process of steps (1) and (2) as NRDC$_{+}^{\prime}$. It is worth noting that for many real-world heterogeneous networks, including the Metabolic network, the critical structural properties (such as degree distribution) exhibit significant self-similarity (or scale invariance) under the aforementioned reduction strategies, as shown in Figures~1(e)(i) and Figure~B1. This indicates that although the scale of the network has been reduced, its inherent structural characteristics have been preserved to some extent.

The subsequent analysis in this paper indicates that for a network with a relatively homogeneous structure, the average degree of the subnetwork typically remains approximately unchanged or tends to decrease after the removal of nodes in step (1). As a result, for this situation, the edge pruning process is not involved in the subnetwork. In contrast, for networks with a more heterogeneous structure (assuming the initial network is connected), the average degree of the subnetworks usually increases following node removal. In this case, it is essential to ensure that during the edge pruning process, the pruned subnetwork maintains a connected structure while keeping the average degree roughly consistent with that of the original network (see Algorithm~\ref{alg:1}).

\begin{algorithm}[htbp]
\caption{Edge pruning algorithm}\label{alg:1}
\begin{algorithmic}[1]
\Require The subgraph $G_i(V, E)$ obtained from step (1), $G_i$ is a connected network, and the average degree $\langle k \rangle_0$ of the initial network $G_0$, the parameters $k_{min}$ and $\delta$.
\Ensure The pruned subgraph $G_i^{\prime}$.
\State $\text{flag} \gets \text{False}$
\While{True}
    \For{each node $u \in V$}
        \State $N_u \gets$ The set of neighboring nodes of the node $u$;
        \If{$|N_v| \leq k_{min}$}
            \State \textbf{continue};
        \EndIf
        \State $N_u^{\prime} \gets$ Sort the nodes in $N_u$ in ascending order according to their degree values;
        \State $e=(u, v) \gets$ Set the node $v = N_u^{\prime}[0]$;
        \If{$e \in E$}
            \State $G_i^{\prime} \gets$ Remove the edge $e$ from $G_i$;
        \EndIf
        \If{$G_i^{\prime}$ is not connected}
            \State Re-add the edge $e$ to $G_i^{\prime}$;
        \EndIf
	     \State $\langle k \rangle_i \gets$ Calculate the average degree of the subgraph $G_i^{\prime}$;
        \If{$\langle k \rangle_i - \langle k \rangle_0 < \delta$}
            \State $\text{flag} \gets \text{True}$; /*In this paper, we set $\delta$ to 0.01*/
            \State \textbf{break};
        \EndIf
        \State $G_i \gets G_i^{\prime}$;
		 \State $E \gets$ Update the edge set of $G_i$;
    \EndFor
    \If{\text{flag}}
        \State \textbf{break};
    \EndIf
\EndWhile
\end{algorithmic}
\end{algorithm}

\textbf{Computational complexity.} Generally speaking, compared to many RG methods, reducing the topological scale of networks from the perspective of subgraph extraction exhibits lower computational complexity~\cite{Chen2024}. For example, when examining the USpowergrid network, which has approximately 5000 nodes, the NRDC$_{+}$ approach can reduce the network size by half in just about 0.05 seconds on the 11th Gen Intel i5-11400 @ 2.60GHz processor. In NRDC$_{+}$, the main consumption lies in calculating DC$_{+}$, and its time complexity is approximately $O(N_0 + M_0)$, where $N_0$ and $M_0$ represent the number of nodes and edges in network $G_0$, respectively. In contrast, the LRG method requires around 70 seconds to achieve the same level of reduction and relies on the complete spectral of the network's Laplacian matrix (the time complexity often reaches $O(N^3)$~\cite{Zhang2025}), which limits its applicability. The GR method consists of two steps: parameter inference and coarse-graining. The running time of parameter inference is approximately 2 to 4 orders of magnitude higher than that of our algorithm. Therefore, reducing complex networks through subgraph extraction offers significant computational advantages and is more suitable for simplifying large-scale complex networks.

\section{Experimental results and discussion}\label{4}
\subsection{Experimental setup}\label{4.1}
\textbf{Implement details.} In this section, we conduct reduction experiments on Erd\"os-R\'enyi (ER) random graphs~\cite{Erdos1959}, Barab\'asi-Albert (BA) scale-free networks~\cite{Barabasi1999}, and twelve real-world networks. The specific research contents are as follows:

$\bullet$ We examine the evolution behavior of the average degree and the relative size of the largest connected component (LCC) during the NRDC$_{+}$ process. 

$\bullet$ We investigate the evolutionary characteristics of Susceptible-Infected-Recovered (SIR)~\cite{Pastor2015} spreading dynamics and information flow~\cite{Zhang2025,De2016,Ghavasieh2024} in synthetic and real-world networks during the NRDC$_{+}$ reduction process. The results show that for most real-world networks, the edge density of subnetworks obtained via NRDC$_{+}$ reduction is generally higher than that of the original networks. Motivated by this finding, we use the NRDC$_{+}^{\prime}$ method to further evaluate the effectiveness of the pruned subnetworks in preserving the spreading dynamics and information flow of the original networks.

$\bullet$ To fully validate the superiority of the NRDC$_{+}^{\prime}$ method, we conduct a comparative analysis with several state-of-the-art network reduction approaches. Finally, this paper discusses the potential application values of the proposed method in practical scenarios and its theoretical implications.

\textbf{Datasets.} The networks investigated in this study include ER random networks, BA scale-free networks, and twelve real-world networks. For ER and BA networks, the number of nodes and the average degree are set to $N = 5000$ and $\langle k \rangle = 10$, respectively. The real-world networks span seven domains: power grids (USpowergrid), biological (Metabolic, Drosophila, and Proteome), language (Music and Words), transportation (Airports), technological (Gnutella and Internet), citation (DBLP), and social networks (Blogs and Enron). The basic topological properties of these networks are summarized in Table~\ref{table:1}. Table~\ref{table:1} reports each network's name, domain, number of nodes, number of edges, average degree, and heterogeneity index~\cite{Hu2008}. In the subsequent experiments, all real-world networks are regarded as undirected and unweighted graphs without self-loops. These real-world network datasets are available from several major open-source repositories$\footnote{\url{http://konect.cc/networks/};}$$\footnote{\url{https://networkrepository.com/};}$$\footnote{\url{https://snap.stanford.edu/data/};}$$\footnote{\url{https://icon.colorado.edu/networks};}$, and the consolidated datasets can be downloaded from url$\footnote{\url{https://github.com/dange-academic/real_network_datasets}.}$.

%============================== Table 1========================
\begin{table}[htbp]
\footnotesize
\centering
\caption{The basic topology information of real-world network datasets.}
\label{table:1}
\tabcolsep 22pt %space between two columns. 用于调整列间距
\begin{tabular*}{\textwidth}{cccccc}
\toprule
Name & Category & $N$ & $M$ & $\langle k\rangle$ & $H$ \\
\hline
Blogs &Social  & 1222 & 16714 & 27.36 & 0.6220 \\
Metabolic & Biological & 1436 & 4718 & 6.57 & 0.4977 \\
Drosophila & Biological & 1770 & 8905 & 10.06 & 0.6506 \\
Music & Language & 2476 & 20624 & 16.66 & 0.7154 \\
Airports & Transportation & 3397 & 19230 & 11.32 & 0.7155 \\
Proteome & Biological & 4100 & 13358 & 6.52 & 0.6690 \\
USpowergrid & Power grids & 4941 & 6594 & 2.67 & 0.3248 \\
Gnutella &Technological  & 6301 & 20777 & 6.59 & 0.5241 \\
Words & Language & 7377 & 44205 & 11.98 & 0.7389 \\
DBLP &Citation  & 12591 & 49635 & 7.88 & 0.6572 \\
Internet & Technological & 23748 & 58414 & 4.92 & 0.6784 \\
Enron & Social & 33696 & 180811 & 10.73 & 0.7285 \\
\bottomrule
\end{tabular*}
\end{table}
%==========================  end Table 1 ======================

\subsection{The evolution behavior of the average degree and LCC of networks during the NRDC$_{+}$ reduction process}\label{4.2}
\begin{figure}[htbp]
\centering
\includegraphics[width=0.6\linewidth]{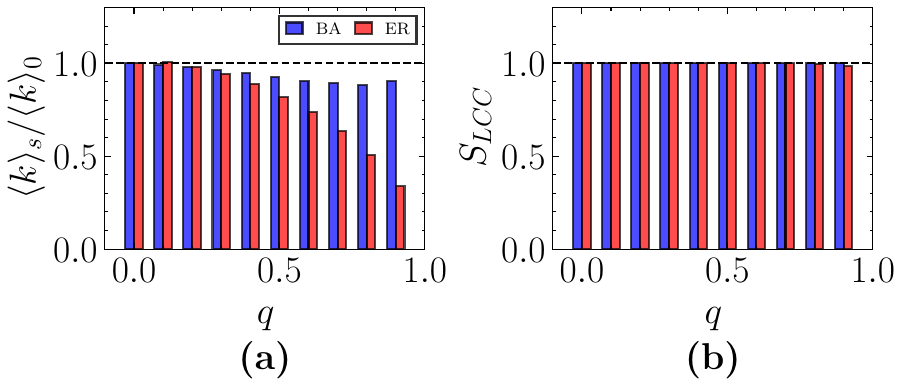}
\caption{The evolution behavior of (a) the relative value of the average degree ($\langle k \rangle_s / \langle k \rangle_0$) and (b) the relative size of the LCC ($S_{LCC}$) in ER and BA synthetic networks during the NRDC$_{+}$ process is analyzed. All results are averaged over 10 independent realizations.}
\label{Fig:2}
\end{figure}

For ER and BA networks, Figure~\ref{Fig:2}(a) explores the relationship between the relative size of the average degree $\langle k \rangle_s / \langle k \rangle_0$ in the residual network (i.e., subnetwork) and the node removal ratio $q$ under the NRDC$_{+}$ process. Here, $\langle k \rangle_s$ denotes the average degree of the residual network, while $\langle k \rangle_0$ represents that of the original network. The results show that during node removal, the average degree of ER random networks exhibits a significant decreasing trend, whereas that of BA scale-free networks remains nearly constant. Furthermore, Figure~\ref{Fig:2}(b) investigates the dependence of the relative size of the LCC $S_{LCC}$ on the node removal ratio $q$, where $S_{LCC} = N_{LCC} / N$ and $N_{LCC}$ is the number of nodes in the LCC. It is observed that BA scale-free networks maintain good connectivity in the node removal process. ER random networks also exhibit relatively good connectivity, becoming disconnected only at higher values of $q$. When the subnetwork is disconnected, the LCC of the subnetwork is treated as a new subgraph for further analysis in subsequent studies.

%==============================Figure 2========================
%============================end Figure 2======================

%==============================Figure 3========================
\begin{figure*}[htbp]
\centering
\includegraphics[width=1.0\linewidth]{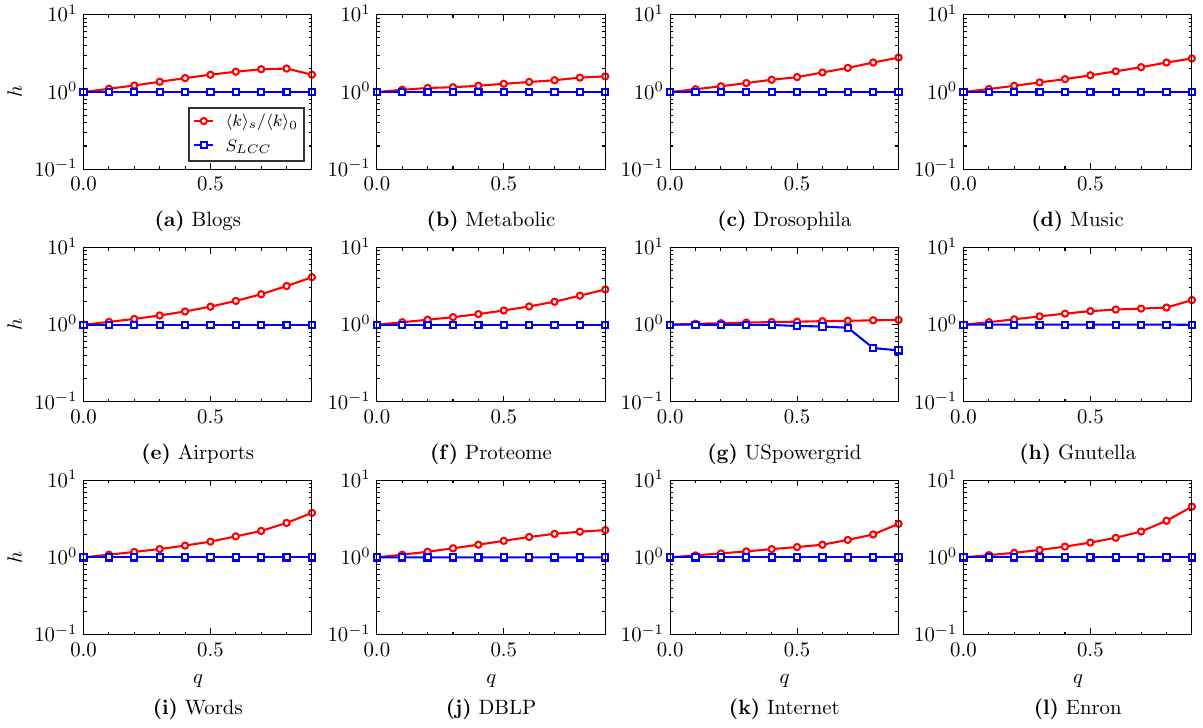}
\caption{Dependence of $\langle k \rangle_s / \langle k \rangle_0$ and $S_{LCC}$ on node removal ratio $q$ ($q \in [0.0,0.9]$) for twelve real-world networks: (a) Blogs, (b) Metabolic, (c) Drosophila, (d) Music, (e) Airports, (f) Proteome, (g) USpowergrid, (h) Gnutella, (i) Words, (j) DBLP, (k) Internet, (l) Enron.}
\label{Fig:3}
\end{figure*}
%============================end Figure 3======================

The red curves in Figure~\ref{Fig:3} show the dependence of the relative average degree $\langle k \rangle_s / \langle k \rangle_0$ on the fraction of removed nodes $q$ for twelve real-world networks (see Table~\ref{table:1}). The results indicate that for the Uspowergrid network, which have relatively low heterogeneity indices $H$ (where the degree distribution exhibits an exponential distribution and lacks significant scale-free properties), the average degree remains nearly constant throughout the NRDC$_{+}$ process. In contrast, the other networks exhibit pronounced scale-free properties (see Figure~B1 in Supplementary File), characterized by high heterogeneity and correspondingly large values of $H$. For these networks, the average degree shows a gradually increasing trend during the same node removal process. The blue curves in Figure~\ref{Fig:3} present the dependence of $S_{LCC}$ on the node removal fraction $q$ for real-world networks. The results show that for the Uspowergrid with weak heterogeneity, $S_{LCC}$ is less than 1 at the larger $q$ value [see Figure~\ref{Fig:3}(g)], indicating that the subnetwork is disconnected. However, for remaining networks, their $S_{LCC}$ remains unchanged and stabilizes around 1, indicating that they consistently maintain good connectivity throughout the NRDC$_{+}$ process.

\subsection{The spreading dynamics of networks during the NRDC$_{+}$ and NRDC$_{+}^{\prime}$ processes}\label{4.3}
We then conduct simulations of SIR epidemic spreading on both the initial networks and their corresponding subnetworks. In the classical SIR model, the population is categorized into three compartments: susceptible (S), infected (I), and recovered (R) states. It is assumed that the total population size is constant, satisfying $N = S + I + R$ (with factors such as birth and death neglected), where $N$ corresponds to the number of nodes in the network. For the initial network $G_0$, the epidemic dynamics are described by the following equations:
\begin{equation}
\begin{cases}
	\displaystyle \frac{dS_0(t)}{dt}=-\beta_0 I_0(t)\frac{S_0(t)}{N_0},\\
	\displaystyle \frac{dI_0(t)}{dt}=\beta_0 I_0(t)\frac{S_0(t)}{N_0}-\gamma_0 I_0(t),\\
	\displaystyle \frac{dR_0(t)}{dt}=\gamma_0 I_0(t),\\
\end{cases}
\end{equation}
where $S_0(t)$, $I_0(t)$, and $R_0(t)$ denote the number of susceptible, infected, and recovered individuals in $G_0$ at time $t$, and $N_0$ is the total number of nodes in $G_0$. Similarly, the epidemic dynamics on a subnetwork $G_l$ are described by the following equations:
\begin{equation}
\begin{cases}
	\displaystyle \frac{dS_l(t)}{dt}=-\beta _l I_l(t)\frac{S_l(t)}{N_l},\\
	\displaystyle \frac{dI_l(t)}{dt}=\beta _l I_l(t)\frac{S_l(t)}{N_l}-\gamma _l I_l(t),\\
	\displaystyle \frac{dR_l(t)}{dt}=\gamma _l I_l(t),\\
\end{cases}
\end{equation}
where $S_l(t)$, $I_l(t)$, and $R_l(t)$ represent the number of susceptible, infected, and recovered nodes in subgraph $G_l$ at time $t$, and $N_l$ is the number of nodes in $G_l$. To facilitate a comparison of spreading dynamics between the initial network and its subnetworks, we assume a consistent infection rate and recovery rate across all simulations, i.e., $\beta_0 = \beta_l = \beta$ and $\gamma_0 = \gamma_l = \gamma$.

In the simulation of SIR spreading dynamics on networks, we consider two scenarios: the first scenario is that the top 10\% of nodes, ranked by their degrees, are selected as the initially infected nodes for both the original network and its corresponding subnetworks, while the remaining nodes are set as susceptible state. In the second scenario, 10\% of nodes are randomly chosen as the initial infected nodes. This paper primarily presents results under the first scenario, and consistent conclusions are also observed in the second scenario. At each time step, infected nodes transmit the infection to their susceptible neighbors at a rate of $\beta$. Subsequently, these infected nodes recover at a rate of $\gamma$. This process continues until there are no infected nodes remaining in the network. For simplicity, the recovery rate is fixed at $\gamma = 1.0$ throughout the study. The subsequent analysis focuses on two key observables: the fraction of infected nodes $i(t) = I(t)/N$ and the fraction of recovered nodes $r(t) = R(t)/N$ at time $t$. These two quantities are direct indicators of the network's epidemic spreading ability. To compute these measures, we employ the EoN$\footnote{\url{https://github.com/springer-math/Mathematics-of-Epidemics-on-Networks}}$(Epidemics on Networks) module, which is a Python package specifically designed to model spreading dynamics such as SIS and SIR processes on networks. To minimize statistical fluctuations, all values of $i(t)$ and $r(t)$ reported in this study are obtained by averaging the results over 100 independent simulation runs for each network and its corresponding subnetwork.

In the subsequent simulations, the node removal ratio is set to $q = 1 - 1/2^l$ ($q$ can also be adjusted to other values between 0 and 1 according to the scenario). This implies that the resulting subnetwork retains $1/2^l$ of the original network's nodes. Here, $l$ is a non-negative integer, where $l = 0$ corresponds to the case with no node removal, i.e., the original network. Notably, since the average degree of both the ER and BA networks did not increase during the NRDC$_{+}$ reduction process (see Figure~\ref{Fig:2}), for these two types of synthetic networks, the NRDC$_{+}^{\prime}$ method is not considered in the subsequent simulation experiments.

\subsubsection{The spreading dynamics of synthetic networks during the NRDC$_{+}$ process}\label{4.3.1}
\begin{figure}[!h]
\centering
\includegraphics[width=1.0\linewidth]{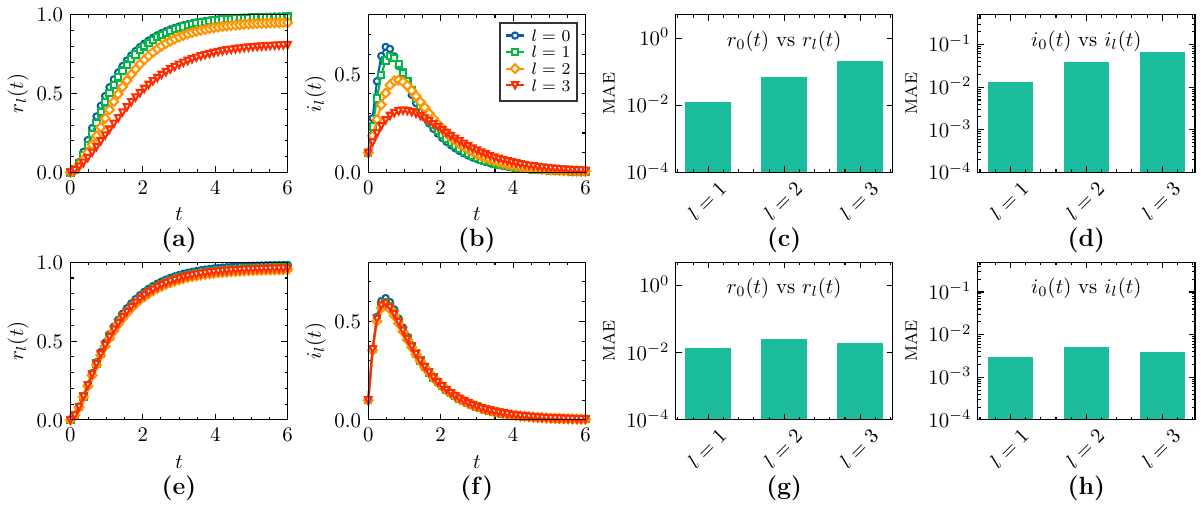}
\caption{The dependence of (a) the proportion of recovered nodes $r(t)$ and (b) the proportion of infected nodes $i(t)$ on the infection time $t$ in the ER random network ($l=0$) and its three subnetworks ($l=1,2,3$), where the subgraphs are obtained via the NRDC$_{+}$ method. (c) and (d) present the mean absolute errors (MAE) between the spreading dynamics curves of the corresponding subgraphs and the original network. (e)-(h) show the results of the BA scale-free network. To eliminate statistical errors, 10 independent sample networks are repeatedly generated as the initial networks for this study. For each initial network and its corresponding subgraph, the results of $i(t)$ and $r(t)$ are presented based on the statistical average of 100 independent realizations.}
\label{Fig:4}
\end{figure}

Figures~\ref{Fig:4}(a) and (b) present the variations of the proportion of recovered nodes $r(t)$ and the proportion of infected nodes $i(t)$ with infection time $t$ for the ER random network ($l = 0$) and its three subgraphs ($l=1, 2, 3$) under the condition that the infection rate $\beta =1.0$. Figures~\ref{Fig:4}(c) and (d) show the mean absolute errors (MAE) between the spreading dynamics curves of the initial network and those of the three subgraphs. The results indicate that as the node removal proportion increases (i.e., as $l$ increases), both $r(t)$ and $i(t)$ of the subgraphs gradually decrease, suggesting that the subgraphs cannot maintain the spreading dynamics of the original ER network. For the BA scale-free network, the results in Figures~\ref{Fig:4}(e)-(h) show that compared with the initial network ($l = 0$), subgraphs of different scales ($l$ = 1, 2, 3) exhibit similar spreading abilities. For example, when $l = 3$, it means that a proportion $q = 1 - 1/2^3 = 0.875$ of nodes are removed from the original BA network, and the resulting subgraph has a node scale only $1/8$ of the original network, yet this subgraph can still reproduce the spreading dynamics of the original BA network well. Based on the comprehensive results of Figure~\ref{Fig:4}, under the NRDC$_{+}$ strategy, among ER and BA networks, only the latter (BA network) shows self-similarity in spreading dynamics.

%==============================Figure 4========================
%============================end Figure 4======================

\begin{figure}[!h]
\begin{center}
\includegraphics[width=1.0\linewidth]{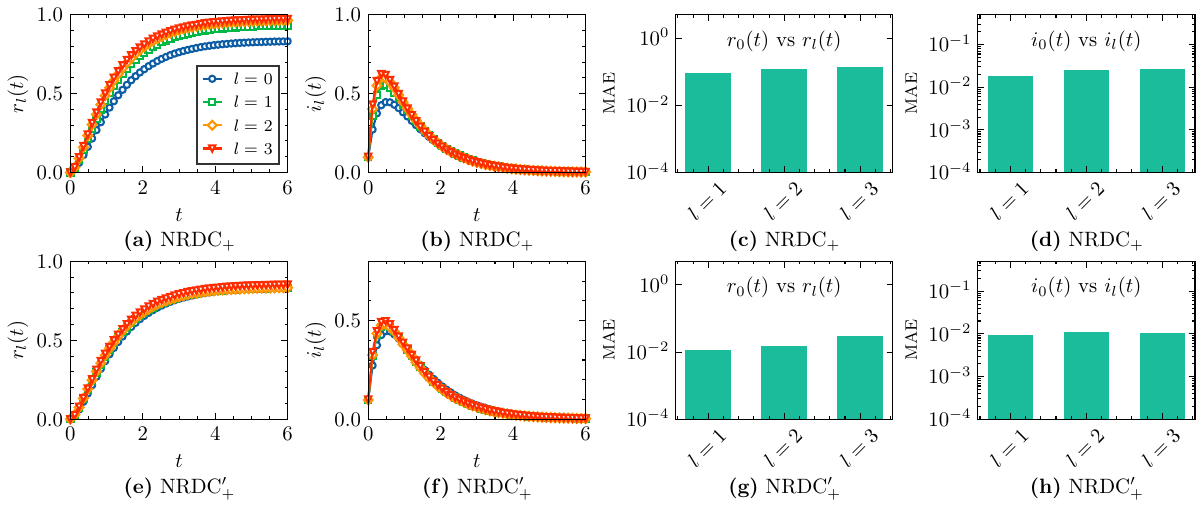}
\caption {The dependence of (a) the proportion of recovered nodes $r(t)$ and (b) the proportion of infected nodes $i(t)$ on the infection time $t$ in the Metabolic network ($l=0$) and its three subnetworks ($l=1,2,3$), where the subnetworks are obtained via the NRDC$_{+}$ method. (c) and (d) show the mean absolute errors (MAE) between the spreading dynamics curves of the corresponding subnetworks and the initial Metabolic network. (e)-(h) show the results of the NRDC$_{+}^{\prime}$ reduction method. To eliminate statistical errors, for the initial Metabolic network and its corresponding subnetworks, the results of $i(t)$ and $r(t)$ are presented based on the statistical average of 100 independent realizations.}
\label{Fig:5}
\end{center}
\end{figure}

\subsubsection{The spreading dynamics of real-world networks during the NRDC$_{+}$ and NRDC$_{+}^{\prime}$ processes}\label{4.3.2}
The results shown in Figure~\ref{Fig:4} suggest that the self-similarity of spreading dynamics during the NRDC$_{+}$ process may depend on two main factors: the topological structure of the network (whether it has a heterogeneous topology) and the evolutionary behavior of the average degree (whether the average degree remains relatively stable). To verify the validity of this speculation, we will conduct experiments similar to Section~\ref{4.3.1} on the real-world networks (see Table~\ref{table:1}). 

Taking the Metabolic network as an example, Figures~\ref{Fig:5}(a) and (b) show the dependence of $r(t)$ and $i(t)$ on the infection time $t$ for this network and its three subnetworks under the infection rate $\beta = 1.0$. The results show that as the proportion of removed nodes increases (i.e., as $l$ increases), the spreading ability of the subnetworks gradually enhances. This trend contrasts with the findings from ER random networks. The results in Figure~\ref{Fig:3} have shown that the average degree of real-world scale-free networks tends to increase with the proportion of removed nodes, which may be the main reason why the spreading ability of subnetworks is significantly stronger than that of the original network. To this end, for subnetworks with an average degree higher than that of the original network, we use the edge pruning algorithm (see Algorithm~\ref{alg:1}) to reduce their average degree to a level similar to that of the original network. Figures~\ref{Fig:5} (e) and (f) present the dependence of $r(t)$ and $i(t)$ on the infection time $t$ in the Metabolic network and its pruned subnetworks. Figures~\ref{Fig:5}(c)(d)(g)(h) show the mean absolute errors (MAE) between the spreading dynamics curves of the initial Metabolic network and its three corresponding subnetworks. The results show that the pruned subnetworks can better reproduce the spreading dynamics of the original Metabolic network. Similar conclusions can be drawn for the other real networks, as shown in Figures~C1-C11.

\begin{figure}[!h]
\begin{center}
\includegraphics[width=0.6\linewidth]{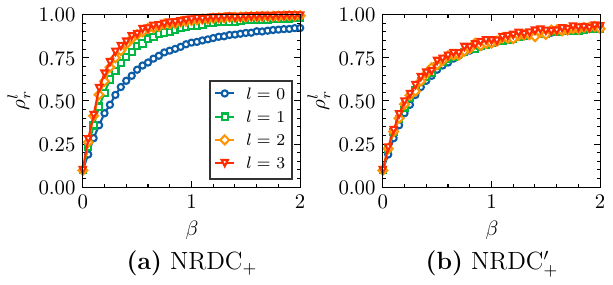}
\caption {The dependence of $\rho_r^l$ on the infection rate $\beta$ in the Metabolic network ($l=0$) and its three subnetworks ($l=1,2,3$). (a) The subnetworks are obtained via the NRDC$_{+}$ method. (b) The subnetworks are obtained via the NRDC$_{+}^{\prime}$ method. To eliminate statistical errors, the result of $\rho_r^l$ for the initial Internet network and each subnetwork is presented based on the statistical average of 100 independent realizations.}
\label{Fig:6}
\end{center}
\end{figure}

To further support these findings, we also studied the results under different infection rates $\beta$. By plotting the dependence of the saturation value $\rho_r = r(t \to \infty)$ of $r(t)$ on the parameter $\beta$, we find that the $\rho_r$ curves still show good self-similarity under the NRDC$_{+}^{\prime}$ method. Fig.~\ref{Fig:6} shows the results of the Metabolic network, and the results for other real-world networks are presented in Figures~C12-C22. For spreading dynamics, this paper only presents the results of the SIR model, when considering other types of epidemic spreading models, such as the SI and SIS models, consistent conclusions can also be drawn.

\subsection{The information flow of networks during the NRDC$_{+}$ and NRDC$_{+}^{\prime}$ processes}\label{4.4}
Next, we investigate the evolutionary behavior of the network's partition function $Z_{\tau}$~\cite{De2016,Ghavasieh2024} under the NRDC$_{+}$ and NRDC$_{+}^{\prime}$ processes. The partition function $Z_{\tau}$ serves as a core concept for understanding information flow in networks, capable of characterizing the propagation characteristics of information at a macroscopic level. Based on the diffusion dynamics of the network, the partition function is defined as:  
\begin{equation}  
Z_{\tau} = \text{Tr}(e^{-\tau \mathbf{L}}),
\end{equation}  
where $\mathbf{L = D - A}$ denotes the Laplacian matrix of the network, $\mathbf{D}$ is a diagonal matrix composed of node degrees, $\mathbf{A}$ is the adjacency matrix, and $\tau$ represents a diffusion scale parameter analogous to time. Essentially, $Z_{\tau}$ embodies the ``total contribution" of all nodes during the diffusion process, reflecting the diversity of information propagation paths and the propagation speed in the network~\cite{Ghavasieh2024}. When $\tau$ is small, the scope of information diffusion is limited, and $Z_{\tau}$ primarily reflects the influence of local structures (such as node degrees and clustering coefficients). When $\tau$ is large, information diffuses globally, and $Z_{\tau}$ captures the macroscopic properties of the entire network (such as the presence of hub nodes and community structures).

The partition function enables a quantitative description of information flow by associating with two critical physical quantities~\cite{Zhang2025}. The first is the network's spectral entropy~\cite{De2016}, defined as:  
\begin{equation}
S_{\tau} = -\text{Tr}(\pmb{\rho}_{\tau} \log \pmb{\rho}_{\tau}),  
\end{equation}  
where $\pmb{\rho}_{\tau} = e^{-\tau \mathbf{L}}/Z_{\tau}$ is the network density matrix. $S_{\tau}$ measures the diversity of information propagation paths: a higher entropy value indicates a greater likelihood of information diffusing through different paths, implying a stronger ``diversion" effect of the network structure on information. The second quantity is the free energy:  
\begin{equation}
F_{\tau} = -\frac{\log Z_{\tau}}{\tau},  
\end{equation}  
which measures the speed of information propagation. A lower free energy signifies faster information diffusion in the network, indicating weaker ``obstruction" of information by the structure. Therefore, the partition function $Z_{\tau}$ encapsulates all the information about entropy and free energy. In this sense, if the $Z_{\tau}$-curves of two networks are similar, their information propagation path diversity and diffusion speed will also be analogous.

\begin{figure}[!h]
\begin{center}
\includegraphics[width=0.6\linewidth]{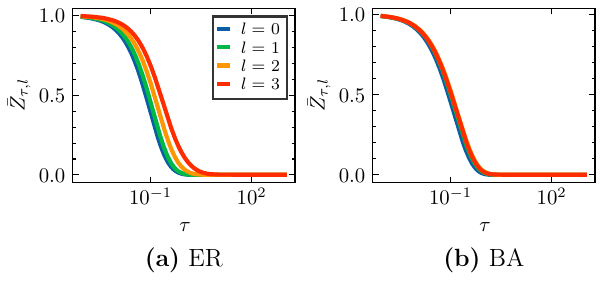}
\caption {The normalized partition function curves of two types of synthetic networks ($l=0$) and three subnetworks ($l$ = 1, 2, 3), where the subgraphs are obtained via the NRDC$_{+}$ method. (a) ER random network. (b) BA scale-free network. To eliminate statistical errors, the statistical averages of 10 independent sample networks are presented here.}
\label{Fig:7}
\end{center}
\end{figure}

Figures~\ref{Fig:7}(a) and (b) show the curves of the normalized partition function $\bar Z_{\tau,l}$ for ER and BA networks, where $\bar Z_{\tau,l} = Z_{\tau,l}/N_l$, and $N_l$ is the number of nodes in subgraph $G_l$. The results reveal significant differences in the partition functions between ER random networks and their corresponding subgraphs. In contrast, the partition functions of BA scale-free networks and their corresponding subgraphs are highly similar. These results indicate that during the NRDC$_{+}$ reduction process, the epidemic dynamics and information flow in ER random networks lack self-similarity, whereas those in BA scale-free networks show notable self-similarity. 

We further investigate the information flow in real-world networks. Our findings indicate that the partition function characterizing network information flow yields results similar to the behavior observed in spreading dynamics. Specifically, the normalized partition function curves for both the pruned subnetworks and the initial networks nearly overlap, these results are shown in Figures~D1-D12.

\textbf{The setting of parameter $k_{min}$.} In pruning Algorithm~\ref{alg:1}, the parameter $k_{min}$ represents the lower bound for the node degree allowed for pruning. Ideally, the optimal value of $k_{min}$ should be determined at the point where the MAE is minimized. As shown in Figure~D13, we plotted the relationship between the MAE values and $k_{min}$ for the Music network, revealing that the optimal value of $k_{min}$ is near 2. We statistically analyzed the optimal $k_{min}$ values for subnetworks at different resolutions, the results indicate that for most networks, the optimal $k_{min}$ values are close to 2. However, a few networks, such as the Metabolic network, has larger optimal $k_{min}$ value, and the optimal value for Metabolic network falls near 10. In this context, for simplicity, $k_{min}$ is set to 2 in this study (unless otherwise specified).

\subsection{Comparison to state-of-the-art methods}\label{4.5}
We apply the NRDC$_{+}$ method and four baseline methods (RDN~\cite{Leskovec2006}, MHRW~\cite{Hubler2008}, CNARW~\cite{Li2019}, MCGS~\cite{Zhao2021}) to the BA scale-free network simultaneously to obtain target subgraphs with the same number of nodes (the number of nodes in the target subgraph is 1/8 of that in the original network). The first row of Figure~E1 shows the spreading dynamics curves of the original BA network and the target subgraphs obtained under different methods. The second row shows the MAE between the spreading dynamics curves of the target subgraphs and the original BA network. Figure~E2 shows the comparison results of the information flow, the results show that the normalized partition function curves of the target subgraph generated by the NRDC$_{+}$ method almost overlap with those of the original BA network, while obvious deviations occur in the other four methods. Overall, at the same level of reduction, the NRDC$_{+}$ method more effectively preserves the spreading dynamics and information flow of the BA scale-free network compared to the other methods.

The results in Figures~\ref{Fig:5}(e)(f) and Figure~\ref{Fig:6}(b) reveal from a macroscopic perspective that the SIR spreading dynamics of subnetworks and the initial network exhibit significant similarity. To accurately quantify how effectively the subnetworks obtained by the NRDC$_{+}^{\prime}$ method reproduce the spreading dynamics of the original network, we will define a measure, denoted as $f_{overlap}$, to characterize the degree of overlap between the spreading dynamics curves of subnetwork $G_l$ and the original network $G_0$. This measure is defined as follows:  
\begin{equation}
f_{overlap}=\frac{1}{1+S_{\Delta}},
\end{equation} 
where $f_{overlap} \in (0, 1]$. A value of $f_{overlap}$ closer to 1 indicates a higher degree of overlap between the two curves. Assuming that $\rho_r$ is a function of the infection rate $\beta$, we denote the spreading dynamics functions for the subnetwork $G_l$ and the original network $G_0$ as $\rho_r^l$ and $\rho_r^0$, respectively. In this study, the range of $\beta$ is set to $[0, 2]$. To calculate $f_{overlap}$, we first use an interpolation function to determine the absolute value of the difference $\Delta$ between $\rho_r^0$ and $\rho_r^l$ at the same $\beta$ points, and then compute the integral $S_{\Delta}$ of $\Delta$ using Simpson's formula. The specific steps for calculating $f_{overlap}$ are as follows: (1) Extract a smaller-scale subgraph $G_l$ from the original network $G_0$ using node removal and edge pruning algorithms, ensuring that the node size of $G_l$ is equal to $1/2^l$ of the size of the original network. (2) Next, calculate the spreading dynamics curves $\rho_r^0$ and $\rho_r^l$ for the original network $G_0$ and subgraph $G_l$, respectively. (3) Finally, apply the interpolation function method (refer to Algorithm~\ref{alg:2} for details) to compute the overlap degree $f_{overlap}$ between these two curves.  

\begin{algorithm}[htbp]
\caption{The algorithm for calculating $f_{overlap}$}\label{alg:2}
\begin{algorithmic}[1]
\Require The spreading ability $\rho_r^0$ and $\rho_r^l$ of the initial network $G_0$ and subgraph $G_l$ at different infection rates $\beta$, where $\beta$, $\rho_r^0$, and $\rho_r^l$ are three one-dimensional arrays of the same length.
\Ensure $f_{overlap}$.
\State Using the interpolating function, two curves $\rho_r^0$ and $\rho_r^l$ are compared at the same $\beta$ point to get $y_0$ = interp1d$(\beta, \rho_r^0)$ and $y_l$ = interp1d$(\beta, \rho_r^l)$;
\State $\beta_{new} \gets$ Reducing the division interval of $\beta$ values, i.e., choosing finer $\beta$ points for calculation;
\State Get the interpolated $\rho_r^0$ and $\rho_r^l$ values, i.e. $\rho_{r,new}^0 \gets y_0(\beta_{new})$, $\rho_{r,new}^l \gets y_l(\beta_{new})$;
\State Calculate the absolute value of the difference between the two curves, $\Delta = |\rho_{r,new}^0 - \rho_{r,new}^l|$;
\State Use Simpson's formula to calculate the integral of $\Delta$, $S_{\Delta} = $ simps$(\Delta, \beta_{new})$;
\State Calculate the degree of overlap $f_{overlap} = 1 / (1 + S_{\Delta})$;
\end{algorithmic}
\end{algorithm}

Table~\ref{table:2} lists the $f_{overlap}$ values for real-world networks under NRDC$_{+}$, NRDC$_{+}^{\prime}$, and four baseline methods, respectively. For ER and BA synthetic networks, only the results under the NRDC$_{+}$ method are considered here (unconsidered results are denoted by ``***"). The second column shows the values of $l$, where $l=3$ implies the node removal ratio is $q=1-1/2^3=0.875$. This means the node size of the resulting subnetwork is only $1/8$ of the original network. To avoid an excessive number of low-degree nodes during the pruning process, the lower bound of the allowable pruning node degree in Algorithm~\ref{alg:1} is set to $k_{min}=2$, as shown in the third column of Table~\ref{table:2}. The fourth to seventh columns display the $f_{overlap}$ values under four baseline sampling methods, where the node sampling rate for each sampling method is set to $sr=1/8$. The eighth column shows the results under NRDC$_{+}$, and the last column presents the results under NRDC$_{+}^{\prime}$. 

%============================== Table 2========================
\begin{table}[htbp]
\footnotesize
\centering
\caption{The value of $f_{overlap}$ is determined by comparing the spreading ability curves $\rho_r^0(\beta)$ of the initial network $G_0$ with $\rho_r^l(\beta)$ of the corresponding subnetwork $G_l$. For four baseline sampling methods (RDN, CNARW, MHRW, and MCGS), the node sampling rate $sr = 1/8$. For Internet and Enron networks, we also considered the results of $l = 4$, $l = 5$, $sr = 1/2^4$, and $sr = 1/2^5$. All the best results are highlighted in bold, and the second-best results are marked with an underline.}
\label{table:2}
\tabcolsep 12pt %space between two columns. 用于调整列间距
\begin{tabular*}{\textwidth}{ccccccccc}
\toprule
Name        &$l$ &$k_{min}$ &RDN &CNARW &MHRW &MCGS &NRDC$_{+}$ &NRDC$_{+}^{\prime}$  \\
\hline
ER($N=5000$)&3 &*** &0.4734  &\underline{0.5954}              &0.5886             &0.5582    &$\mathbf{0.7011}$  &*** \\
BA($N=5000$)&3 &*** &0.6812  &0.7666              &0.6395             &\underline{0.7950}    &$\mathbf{0.9527}$  &*** \\
Blogs   &3 &2   &0.8446  &0.8634   &0.8940             &\underline{0.9593}    &0.7982             &$\mathbf{0.9657}$ \\
Metabolic   &3 &2   &0.9450  &$\mathbf{0.9834}$   &0.9209             &0.9268    &0.7794             &\underline{0.9489} \\
Drosophila  &3 &2   &0.8199  &0.8058              &$\mathbf{0.9530}$  &\underline{0.9338}    &0.6982             &0.9142 \\
Music       &3 &2   &0.8016  &0.8261              &0.7961             &\underline{0.8817}    &0.7448             &$\mathbf{0.9620}$ \\
Airports    &3 &2   &0.7284  &0.7095              &\underline{0.8836}             &0.7479    &0.6282             &$\mathbf{0.9551}$ \\
Proteome    &3 &2   &0.7691  &0.7796              &0.8910             &\underline{0.9102}    &0.6450             &$\mathbf{0.9165}$ \\
USpowergrid &3 &2   &0.8943  &\underline{0.9218}              &$\mathbf{0.9391}$             &0.8057    &0.8331  &0.9210 \\
Gnutella    &3 &2   &0.7579  &\underline{0.8720}              &0.8248             &0.7763    &0.7610  &$\mathbf{0.9221}$ \\
Words       &3 &2   &0.8021  &0.7946              &0.8238             &\underline{0.8633}    &0.7003             &$\mathbf{0.9616}$ \\
DBLP        &3 &2   &0.7851  &0.7874              &\underline{0.8534} &0.7897    &0.6706             &$\mathbf{0.9304}$ \\
Internet    &3 &2   &0.8121  &\underline{0.8610}              &0.8397             &0.7865    &0.6972             &$\mathbf{0.9143}$ \\
Enron       &3 &2   &\underline{0.7778}  &0.7587              &0.7646             &0.7673    &0.6509             &$\mathbf{0.9711}$ \\
Internet    &4 &2   &0.8092  &0.8445              &\underline{0.8699}             &0.7736    &0.6553             &$\mathbf{0.8997}$ \\
Enron       &4 &2   &0.7682  &0.7514              &\underline{0.8229}             &0.7379    &0.6357             &$\mathbf{0.9771}$ \\
Internet    &5 &2   &0.8058  &0.8654              &\underline{0.9096}             &0.7243    &0.6224             &$\mathbf{0.9164}$ \\
Enron       &5 &2   &0.7917  &0.7369              &\underline{0.8727}             &0.7257    &0.6323             &$\mathbf{0.9781}$ \\
\bottomrule
\end{tabular*}
\end{table}
%==========================  end Table 2 ======================

Our results show that the NRDC$_{+}^{\prime}$ method shows strong competitiveness compared to the other four baseline methods, as highlighted in bold in the last column of Table~\ref{table:2}, and almost all $f_{overlap}$ values exceed 0.9. This result further demonstrates that we can accurately predict the spreading dynamics behavior of the original network based on the reduced subgraph. For instance, in the Enron network, when $l=5$, the node size of the subnetwork is only $1/32$ of the initial network. At this time, the average prediction accuracy for the spreading ability of the original network—based on the subgraph obtained through the NRDC$_{+}^{\prime}$ algorithm—reaches as high as 97\%. Furthermore, even with a moderate degree of reduction, such as setting $l$ to 1 and 2 for the NRDC$_{+}^{\prime}$ methods, and using node sampling rates $sr$ of 1/2 and 1/4 for the other four baseline sampling methods, the NRDC$_{+}^{\prime}$ method still displays strong competitiveness, as shown in Tables~E1 and E2. We further investigate the ability of the NRDC$_{+}^{\prime}$ method to maintain information flow in real-world networks. Our findings indicate that the partition function characterizing network information flow yields results similar to those observed in spreading dynamics.

As a supplement, we also compared the NRDC$_{+}^{\prime}$ method with several mainstream renormalization approaches, including the GR~\cite{Garcia2018} (and its pruned version GR$^{\prime}$), LRG~\cite{Villegas2023}, and LCG~\cite{loures2023}. The results from both spreading dynamics and partition function analyses demonstrate that the NRDC$_{+}^{\prime}$ method significantly outperforms other methods, as shown in Figures~E3-E10. Considering that GR and LCG have excessively high computational costs, we only present the results for the Blogs, Metabolic, Drosophila, and Music networks, which have relatively small numbers of nodes. Overall, our method is a subgraph extraction strategy based on DC$_{+}$ and is integrated with an edge pruning process. Technically, it is straightforward to implement, resulting in lower computational complexity compared to renormalization methods (see Figure~E11).

\subsection{Potential applications and significance}\label{4.6}
\begin{figure}[!h]
\begin{center}
\includegraphics[width=0.6\linewidth]{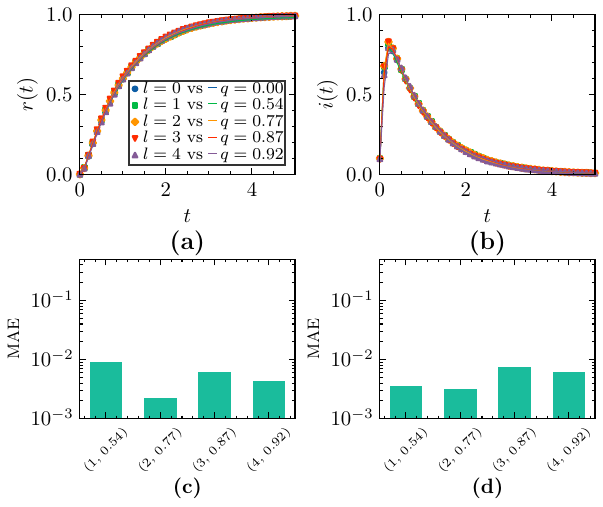}
\caption {Comparison of spreading dynamics in the empirical multiscale human connectome networks (symbols) and the predicted spreading dynamics by the NRDC$_{+}^{\prime}$ method (solid lines). (a) $r(t)$ vs $t$, (b) $i(t)$ vs $t$. (c) and (d) show the mean absolute errors (MAE) between the spreading dynamics curve of the $l$-layer ($l=1,2,3,4$) empirical human connectome network and the spreading dynamics curve of the subnetwork obtained by the NRDC$_{+}^{\prime}$ method. In the NRDC$_{+}^{\prime}$ method, the node removal ratios are set as $q =$ 0.54, 0.77, 0.87 and 0.92 to ensure that the number of nodes of these subnetworks match those of the empirical human connectomes labeled $l=1,2,3,4$, respectively.}
\label{Fig:8}
\end{center}
\end{figure}

\begin{figure}[!h]
\begin{center}
\includegraphics[width=0.6\linewidth]{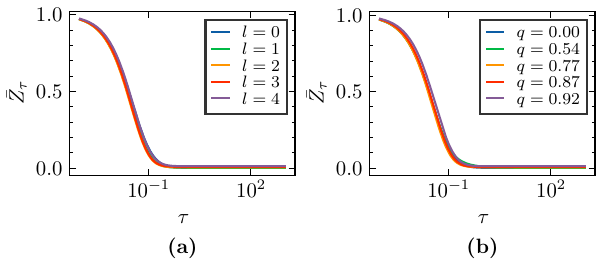}
\caption {(a) The normalized partition functions of multiscale human connectome networks with five anatomical resolutions. (b) The normalized partition functions of the $l=0$ layer human connectome network and four gradually shrinking subnetworks derived by NRDC$_{+}^{\prime}$ method.}
\label{Fig:9}
\end{center}
\end{figure}

Exploring the self-similarity of networks during the reduction process is of significant value for predicting their critical dynamic characteristics. For instance, we apply the NRDC$_{+}^{\prime}$ method to predict the spreading dynamics and information flow of the human connectome networks (at five anatomical resolutions~\cite{Zheng2020geometric}, labeled as $l=0,1,2,3,4$). These resolutions correspond to approximately 1014, 462, 233, 128, and 82 nodes, respectively. Our goal is to determine whether the method could reproduce empirical observation results. We first investigate the spreading dynamics of these five connectome networks at different resolutions. The results show that the networks at different scales exhibit highly similar spreading dynamics behaviors, as indicated by the symbols in Figures~\ref{Fig:8}(a) and (b). Next, we applied the NRDC$_{+}^{\prime}$ method to the highest-resolution network ($l=0$) to obtain four subnetworks with the same number of nodes as the networks labeled $l=1,2,3,4$. Notably, the predictions from the NRDC$_{+}^{\prime}$ method show a striking consistency with empirical observations, as illustrated by the solid lines in Figures~\ref{Fig:8}(a) and (b). Additionally, similar phenomena are also observed for the partition function, as shown in Figure~\ref{Fig:9}.

In future studies, the findings of this study can also give rise to numerous applications. Here, we present three possible application examples:

\textbf{Rapid evaluation of infectious disease prevention and control strategies~\cite{Ritter2017}.} In reality, urban population contact networks (such as interpersonal interaction networks formed by communities, transportation networks, and workplaces) are extremely large-scale, with the number of nodes reaching millions or even tens of millions. Directly simulating disease transmission on the original network to evaluate the effectiveness of prevention and control measures (e.g., lockdown scope, vaccination rate) often faces issues of high computational cost and slow iteration speed. By leveraging the self-similarity of reduced networks, the original network can be simplified into smaller ``miniature networks'' that retain consistent transmission characteristics. Through rapid simulation of key indicators (such as infection peaks, duration, and final infection scale) under different prevention and control strategies on these miniature networks, it becomes feasible to efficiently screen out optimal strategies, providing decision support for public health authorities.  

\textbf{Optimization of interventions in social media information dissemination~\cite{Xu2019Inferring,Notarmuzi2022}.} Social media networks (e.g., Weibo, Twitter) consist of billions of user nodes, and the dissemination paths of information (such as rumors and hot events) are complex. If intervention in information diffusion is required (e.g., restricting the dissemination capacity of key nodes), directly simulating the dissemination process on the original network incurs extremely high costs. If reduced networks can preserve the core dynamic characteristics of information dissemination (e.g., transmission thresholds, the role of key opinion leaders), relevant simulations can be conducted on these reduced networks. This enables rapid identification of ``dissemination hubs'' (e.g., high-influence accounts), prediction of the ultimate scope of information diffusion, and evaluation of the effectiveness of intervention measures (such as ``limiting the flow of specific nodes'' and ``releasing clarifying information''), thereby assisting platforms in efficiently managing information dissemination.

\textbf{Study on the neurological diseases spreading and the stability of functional modules in brain functional networks~\cite{Sendi2025}.} The functionality of brain networks relies on Laplacian spectral properties and partition functions, both of which reflect the synchronization of functional modules in brain regions and the efficiency of information integration. Reduced networks can quickly identify the ``core hubs'' of functional modules through spectral properties while retaining these two characteristics (e.g., the spectrally sensitive nodes of the hippocampus in memory networks), so as to analyze the impact of disease spread on module stability.

\section{Conclusion}\label{5}
This paper proposes a network reduction algorithm based on the enhanced degree centrality (DC$_{+}$) metric, aimed at evaluating its effectiveness in preserving the spreading dynamics and information flow of real networks. The core idea of the algorithm is to preferentially remove nodes with lower DC$_{+}$ values from the network to obtain a smaller subgraph. Furthermore, when the density of the subgraph is higher than that of the original network, an edge pruning algorithm is proposed to reduce the subgraph's density to a level close to that of the original network. Experimental results on synthetic and real-world networks reveal the following findings: the spreading dynamics and partition function of ER random networks do not exhibit self-similarity under the proposed reduction algorithm, whereas BA scale-free networks show strong self-similarity. For real scale-free networks with obvious heterogeneous topological structures, their average degree gradually increases during node removal, and the spreading ability typically correlates positively with the average degree. For such networks, the results show that when the pruning algorithm is used to adjust the average degree of the subgraph to a level close to that of the original network, the spreading dynamics and information flow behavior of the original network can be effectively reproduced.

In essence, the network reduction scheme proposed in this paper differs significantly from other renormalization techniques that rely on coarse-graining principles. This scheme is not only technically simpler but also more efficient in execution compared to traditional methods. In addition, by introducing a parameter for the node removal ratio, the network can be effectively scaled down to a more manageable size, facilitating analysis and modeling. The findings of this study indicate that within the deeper structures of large networks, there may exist one or more smaller subgraphs whose spreading dynamics are extremely similar to those of the original network. This insight provides a potential solution for simplifying the topological structure of large-scale networks. From a practical perspective, network reduction provides a fast estimation method. It transforms the dynamic modeling traditionally conducted on an initial large-scale network into a more streamlined process on a smaller network, which is of great practical significance for predicting the dynamic behavior of large networks and accelerating model simulation.

%%%%%%%%%%%%%%%%%%%%%%%%%%%%%%%%%%%%%%%%%%%%%%%%%%%%%%%
%%% Acknowledgements. 致谢
%%%%%%%%%%%%%%%%%%%%%%%%%%%%%%%%%%%%%%%%%%%%%%%%%%%%%%%
\section*{Acknowledgements}
This work was supported in part by the National Natural Science Foundation of China under Grant Nos.~62425602 and 62273159, and in part by the Natural Science Foundation of Hubei Province of China under Grant Nos.~2025AFA027 and 2025AFB234.

%%%%%%%%%%%%%%%%%%%%%%%%%%%%%%%%%%%%%%%%%%%%%%%%%%%%%%%
%%% Supplements. 补充材料, 非必选
%%%%%%%%%%%%%%%%%%%%%%%%%%%%%%%%%%%%%%%%%%%%%%%%%%%%%%%
\section*{Supplementary File}
The Appendixes A-E present additional experimental results on the spreading dynamics and partition functions of BA scale-free networks and some real-world networks. We also provide comparison results between our method and other network reduction approaches.

\bibliographystyle{unsrt}  
\bibliography{references}  %%% Remove comment to use the external .bib file (using bibtex).
%%% and comment out the ``thebibliography'' section.

\setcounter{figure}{0}
\renewcommand{\thefigure}{A\arabic{figure}}  % 编号为"A1", "A2"等

\section*{Appendix}

\begin{appendix}

\setcounter{figure}{0}
\renewcommand{\thefigure}{A\arabic{figure}}  % 编号为"A1", "A2"等

\section{The DC and DC$_{+}$ of real-world networks}
Figure~A1 presents the degree centrality (DC) values of all nodes in ten real-world networks, with the data arranged in ascending order. It is observed that there are a large number of nodes with the same degree value in the networks, a characteristic visually reflected by the horizontal segments in the figure, indicating that traditional DC fails to effectively distinguish differences among such nodes. To address this limitation, we propose an enhanced degree centrality metric, DC$_{+}$, which is defined as the product of a node's degree and its average neighbor degree. The results in Figure~A2 demonstrate that DC$_{+}$ significantly improves the ability to discriminate differences between nodes.

%==============================Figure S1========================
\begin{figure*}[htbp]
\centering
\includegraphics[width=1.0\linewidth]{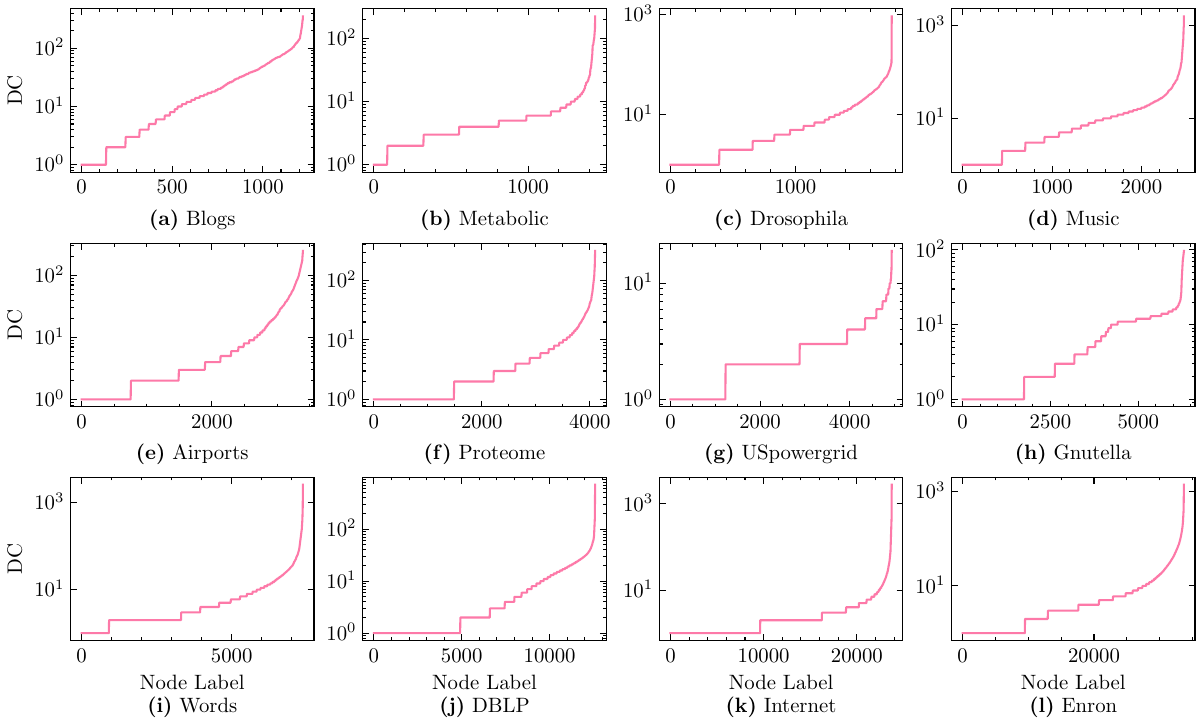}
\caption{The DC values of real-world networks. The horizontal axis represents the labels of the nodes, and the vertical axis represents the degree values of the nodes. (a) Blogs, (b) Metabolic, (c) Drosophila, (d) Music, (e) Airports, (f) Proteome, (g) USpowergrid, (h) Gnutella, (i) Words, (j) DBLP, (k) Internet, (l) Enron.}
\label{Fig:S1}
\end{figure*}
%============================end Figure S1======================

%==============================Figure S2========================
\begin{figure*}[htbp]
\centering
\includegraphics[width=1.0\linewidth]{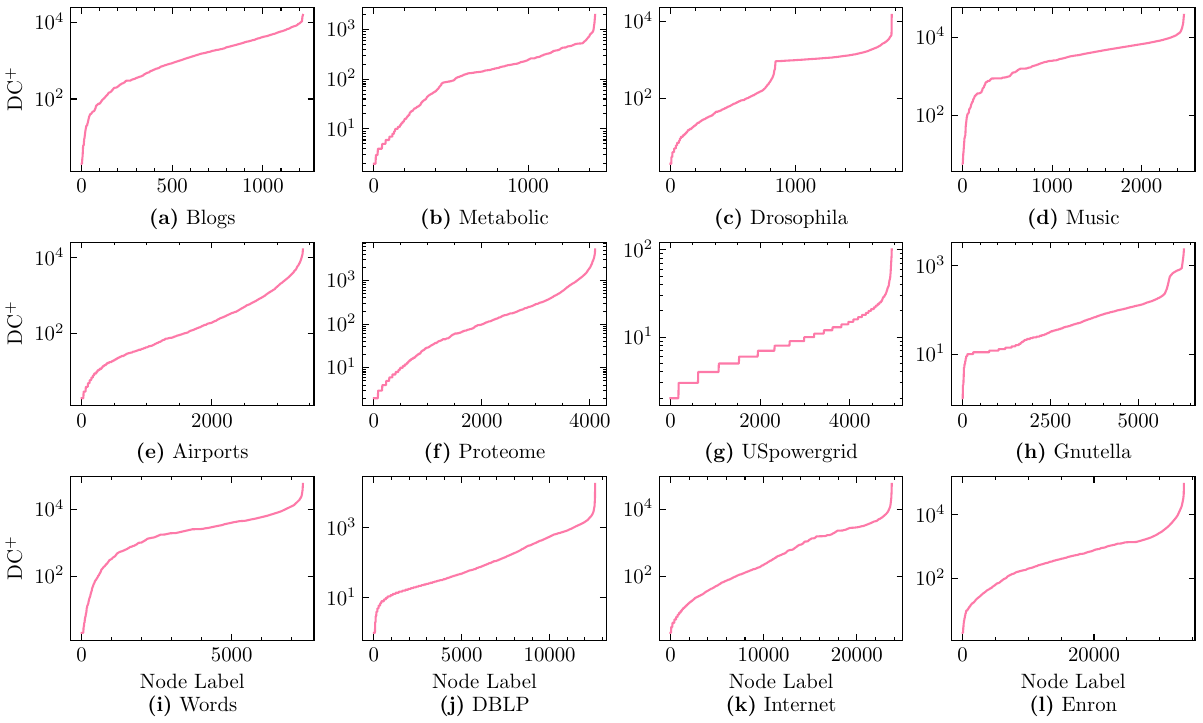}
\caption{The DC$_{+}$ values of real-world networks. The horizontal axis represents the labels of the nodes, and the vertical axis represents the DC$_{+}$ values of the nodes. (a) Blogs, (b) Metabolic, (c) Drosophila, (d) Music, (e) Airports, (f) Proteome, (g) USpowergrid, (h) Gnutella, (i) Words, (j) DBLP, (k) Internet, (l) Enron.}
\label{Fig:S2}
\end{figure*}
%============================end Figure S2======================

\clearpage

\setcounter{figure}{0}
\renewcommand{\thefigure}{B\arabic{figure}}

\section{The degree distributions of real-world networks during the NRDC$_{+}^{\prime}$ reduction process}
The results in Figure B1 demonstrate that the complementary cumulative degree distribution (CCDF) of the network exhibits approximate self-similarity during the NRDC$_{+}^{\prime}$ reduction process. The insets show the information entropy $S$ of the degree distribution, which remains nearly constant as $l$ increases (i.e., as the node removal ratio increases). This further indicates that the NRDC$_{+}^{\prime}$ strategy largely preserves the degree distribution characteristics of the original network.  

%==============================Figure S3========================
\begin{figure*}[htbp]
\centering
\includegraphics[width=1.0\linewidth]{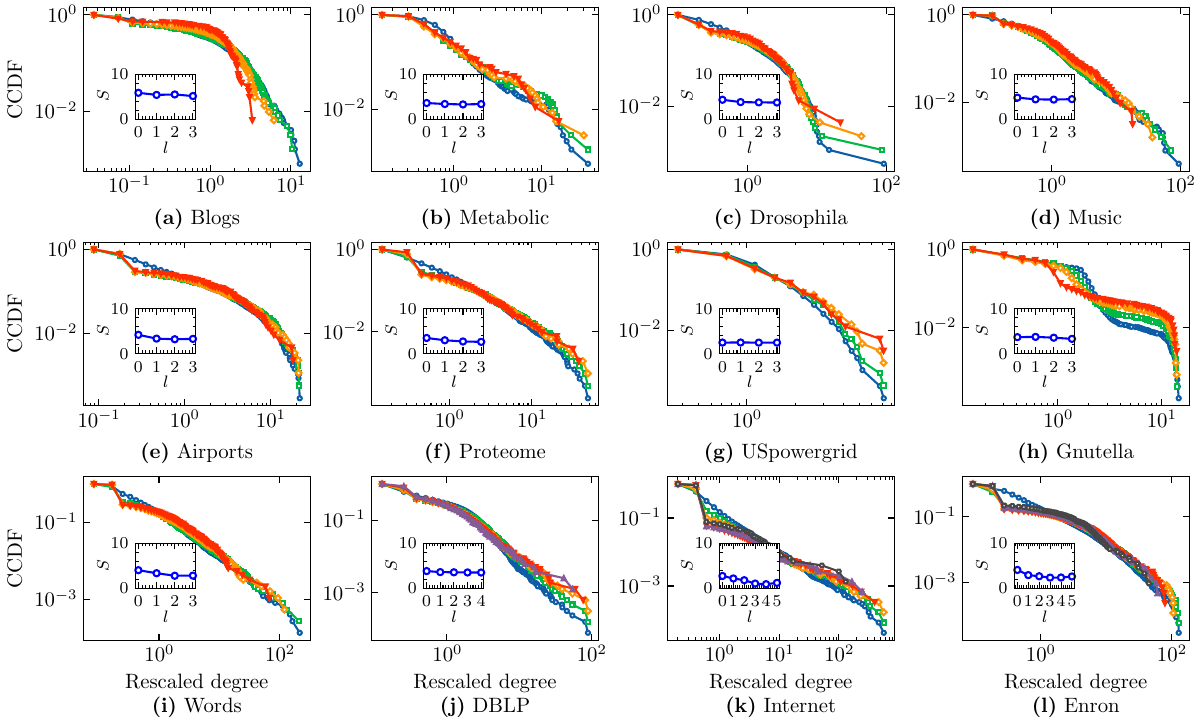}
\caption{Self-similarity of the complementary cumulative degree distribution (CCDF) of real-world networks during the NRDC$_{+}^{\prime}$ reduction process, where the abscissa represents the rescaled degree values. (a) Blogs, (b) Metabolic, (c) Drosophila, (d) Music, (e) Airports, (f) Proteome, (g) USpowergrid, (h) Gnutella, (i) Words, (j) DBLP, (k) Internet, (l) Enron.}
\label{Fig:S3}
\end{figure*}
%============================end Figure S2======================

\newpage

\setcounter{figure}{0}
\renewcommand{\thefigure}{C\arabic{figure}}

\section{The spreading dynamics of real-world networks during the NRDC$_{+}$ and NRDC$_{+}^{\prime}$ processes}
In addition to the spreading dynamics of the Metabolic network presented in the main text (see Figures 5 and 6), we conduct similar investigations on other real-world networks (see Figures C1-C22). The results demonstrate that subnetworks obtained by the NRDC$_{+}^{\prime}$ method can better reproduce the spreading dynamics observed in original networks. In all simulation experiments, to eliminate statistical errors, for the initial network and its corresponding subnetworks, the results of $i(t)$ and $r(t)$ are presented based on the statistical average of 100 independent realizations.

\begin{figure}[htbp]
\begin{center}
\includegraphics[width=1.0\linewidth]{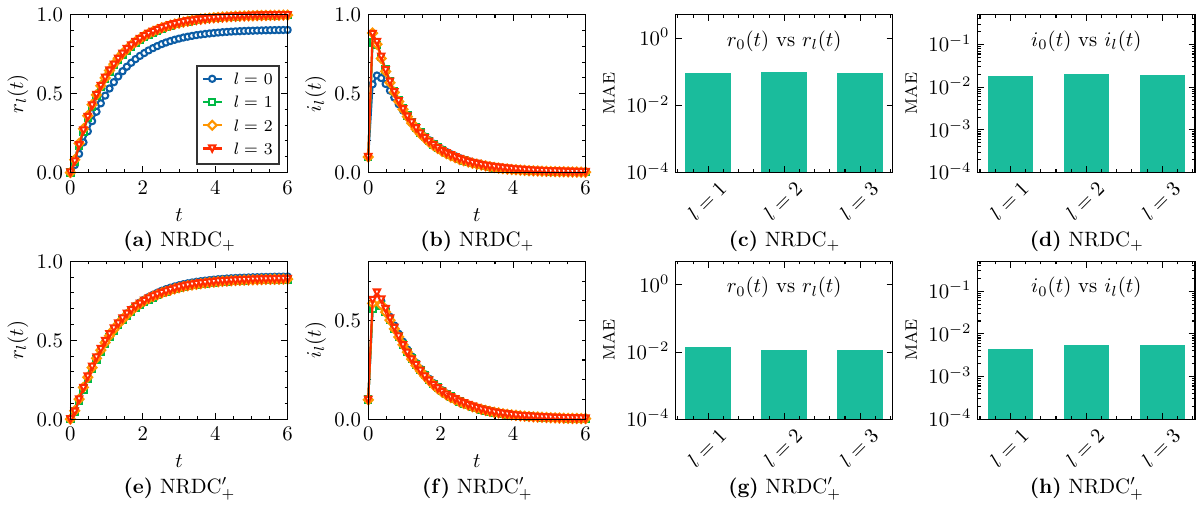}
\caption {The dependence of (a) the proportion of recovered nodes $r(t)$ and (b) the proportion of infected nodes $i(t)$ on the infection time $t$ in the Blogs network ($l=0$) and its three subnetworks ($l=1,2,3$), where the subnetworks are obtained via the NRDC$_{+}$ method. (c) and (d) show the mean absolute errors (MAE) between the epidemic dynamics curves of the corresponding subnetworks and the initial Blogs network. (e)-(h) show the results of the NRDC$_{+}^{\prime}$ reduction method.}
\label{Fig:S4}
\end{center}
\end{figure}

\begin{figure}[htbp]
\begin{center}
\includegraphics[width=1.0\linewidth]{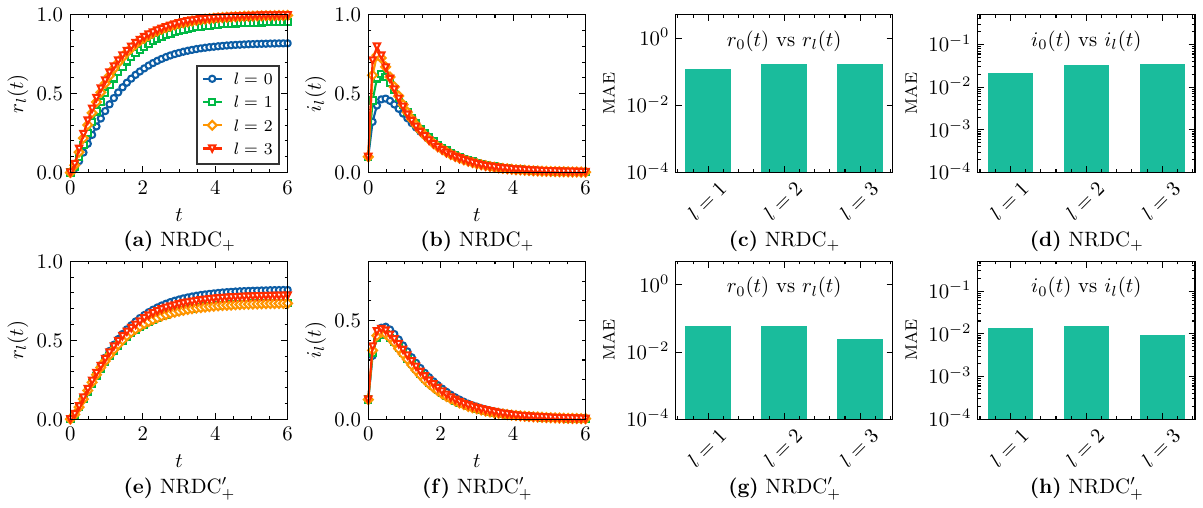}
\caption {The dependence of (a) the proportion of recovered nodes $r(t)$ and (b) the proportion of infected nodes $i(t)$ on the infection time $t$ in the Drosophila network ($l=0$) and its three subnetworks ($l=1,2,3$), where the subnetworks are obtained via the NRDC$_{+}$ method. (c) and (d) show the mean absolute errors (MAE) between the epidemic dynamics curves of the corresponding subnetworks and the initial Drosophila network. (e)-(h) show the results of the NRDC$_{+}^{\prime}$ reduction method.}
\label{Fig:S5}
\end{center}
\end{figure}

\begin{figure}[htbp]
\begin{center}
\includegraphics[width=1.0\linewidth]{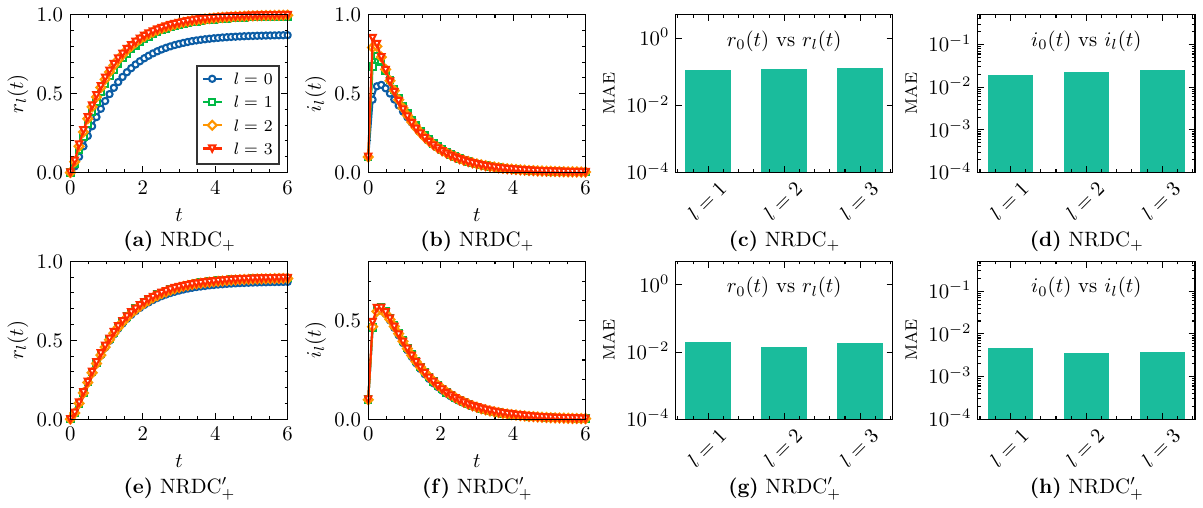}
\caption {The dependence of (a) the proportion of recovered nodes $r(t)$ and (b) the proportion of infected nodes $i(t)$ on the infection time $t$ in the Music network ($l=0$) and its three subnetworks ($l=1,2,3$), where the subnetworks are obtained via the NRDC$_{+}$ method. (c) and (d) show the mean absolute errors (MAE) between the epidemic dynamics curves of the corresponding subnetworks and the initial Music network. (e)-(h) show the results of the NRDC$_{+}^{\prime}$ reduction method.}
\label{Fig:S6}
\end{center}
\end{figure}

\begin{figure}[htbp]
\begin{center}
\includegraphics[width=1.0\linewidth]{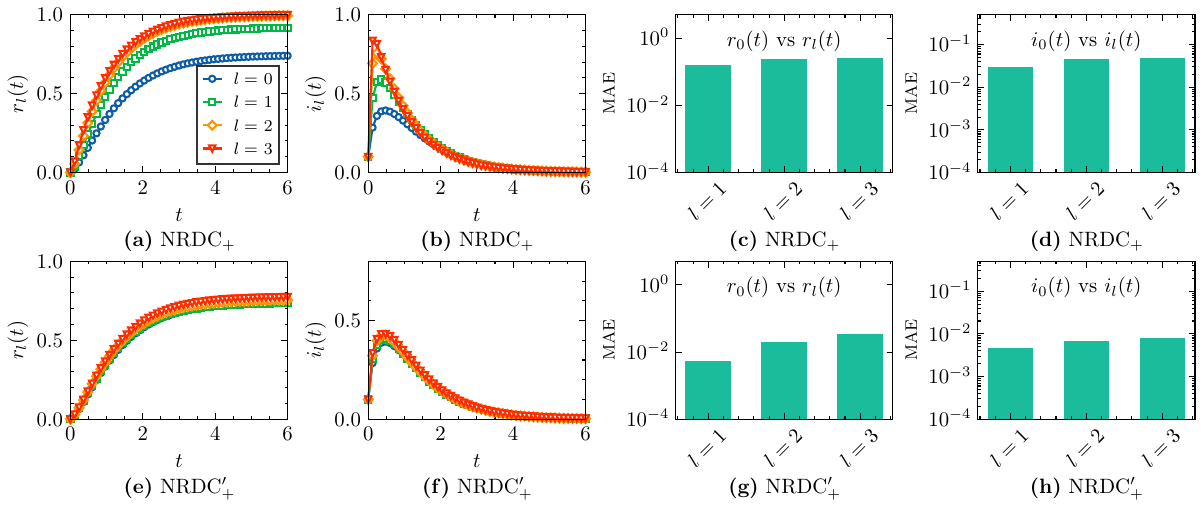}
\caption {The dependence of (a) the proportion of recovered nodes $r(t)$ and (b) the proportion of infected nodes $i(t)$ on the infection time $t$ in the Airports network ($l=0$) and its three subnetworks ($l=1,2,3$), where the subnetworks are obtained via the NRDC$_{+}$ method. (c) and (d) show the mean absolute errors (MAE) between the epidemic dynamics curves of the corresponding subnetworks and the initial Airports network. (e)-(h) show the results of the NRDC$_{+}^{\prime}$ reduction method.}
\label{Fig:S7}
\end{center}
\end{figure}

\begin{figure}[htbp]
\begin{center}
\includegraphics[width=1.0\linewidth]{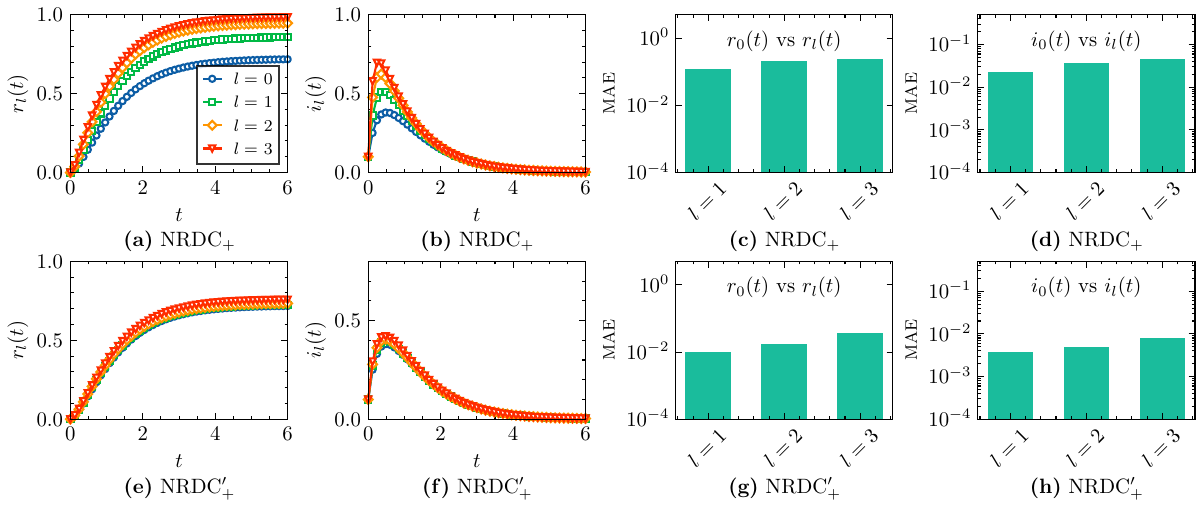}
\caption {The dependence of (a) the proportion of recovered nodes $r(t)$ and (b) the proportion of infected nodes $i(t)$ on the infection time $t$ in the Proteome network ($l=0$) and its three subnetworks ($l=1,2,3$), where the subnetworks are obtained via the NRDC$_{+}$ method. (c) and (d) show the mean absolute errors (MAE) between the epidemic dynamics curves of the corresponding subnetworks and the initial Proteome network. (e)-(h) show the results of the NRDC$_{+}^{\prime}$ reduction method.}
\label{Fig:S8}
\end{center}
\end{figure}

\begin{figure}[htbp]
\begin{center}
\includegraphics[width=1.0\linewidth]{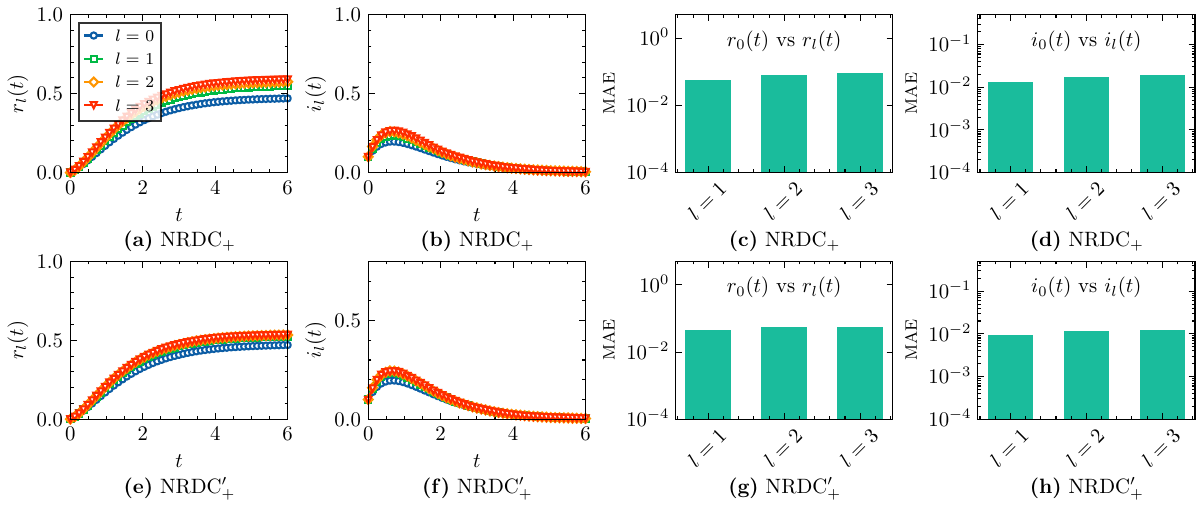}
\caption {The dependence of (a) the proportion of recovered nodes $r(t)$ and (b) the proportion of infected nodes $i(t)$ on the infection time $t$ in the USpowergrid network ($l=0$) and its three subnetworks ($l=1,2,3$), where the subnetworks are obtained via the NRDC$_{+}$ method. (c) and (d) show the mean absolute errors (MAE) between the epidemic dynamics curves of the corresponding subnetworks and the initial USpowergrid network. (e)-(h) show the results of the NRDC$_{+}^{\prime}$ reduction method.}
\label{Fig:S9}
\end{center}
\end{figure}

\begin{figure}[htbp]
\begin{center}
\includegraphics[width=1.0\linewidth]{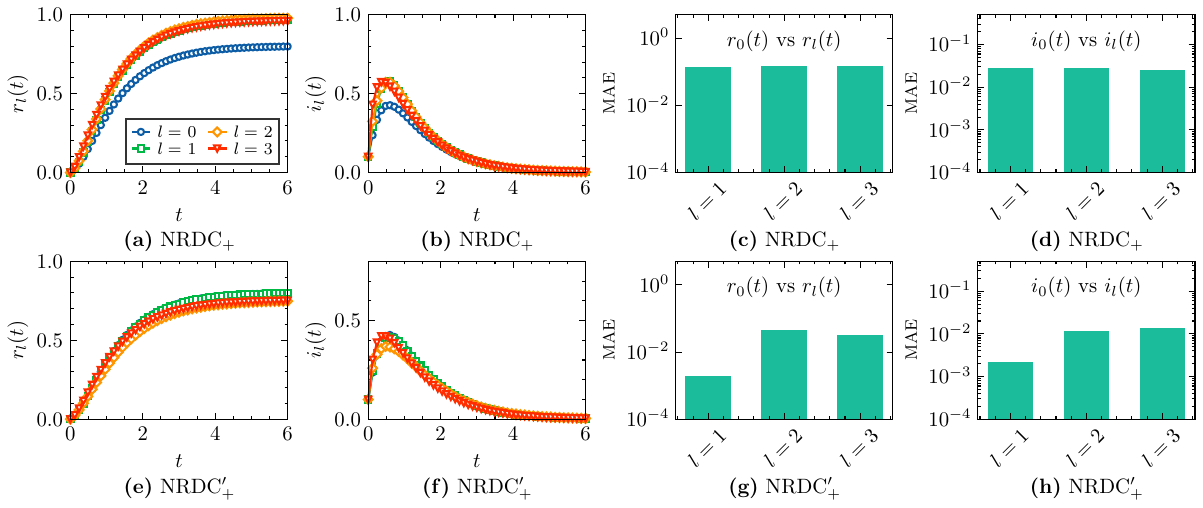}
\caption {The dependence of (a) the proportion of recovered nodes $r(t)$ and (b) the proportion of infected nodes $i(t)$ on the infection time $t$ in the Gnutella network ($l=0$) and its three subnetworks ($l=1,2,3$), where the subnetworks are obtained via the NRDC$_{+}$ method. (c) and (d) show the mean absolute errors (MAE) between the epidemic dynamics curves of the corresponding subnetworks and the initial Gnutella network. (e)-(h) show the results of the NRDC$_{+}^{\prime}$ reduction method.}
\label{Fig:S10}
\end{center}
\end{figure}

\begin{figure}[htbp]
\begin{center}
\includegraphics[width=1.0\linewidth]{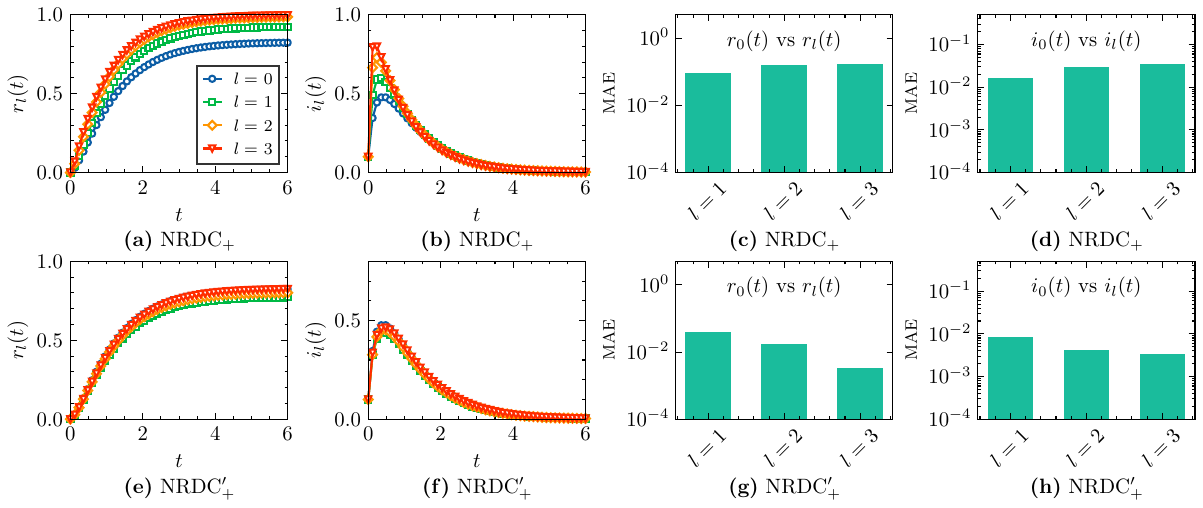}
\caption {The dependence of (a) the proportion of recovered nodes $r(t)$ and (b) the proportion of infected nodes $i(t)$ on the infection time $t$ in the Words network ($l=0$) and its three subnetworks ($l=1,2,3$), where the subnetworks are obtained via the NRDC$_{+}$ method. (c) and (d) show the mean absolute errors (MAE) between the epidemic dynamics curves of the corresponding subnetworks and the initial Words network. (e)-(h) show the results of the NRDC$_{+}^{\prime}$ reduction method.}
\label{Fig:S11}
\end{center}
\end{figure}

\begin{figure}[htbp]
\begin{center}
\includegraphics[width=1.0\linewidth]{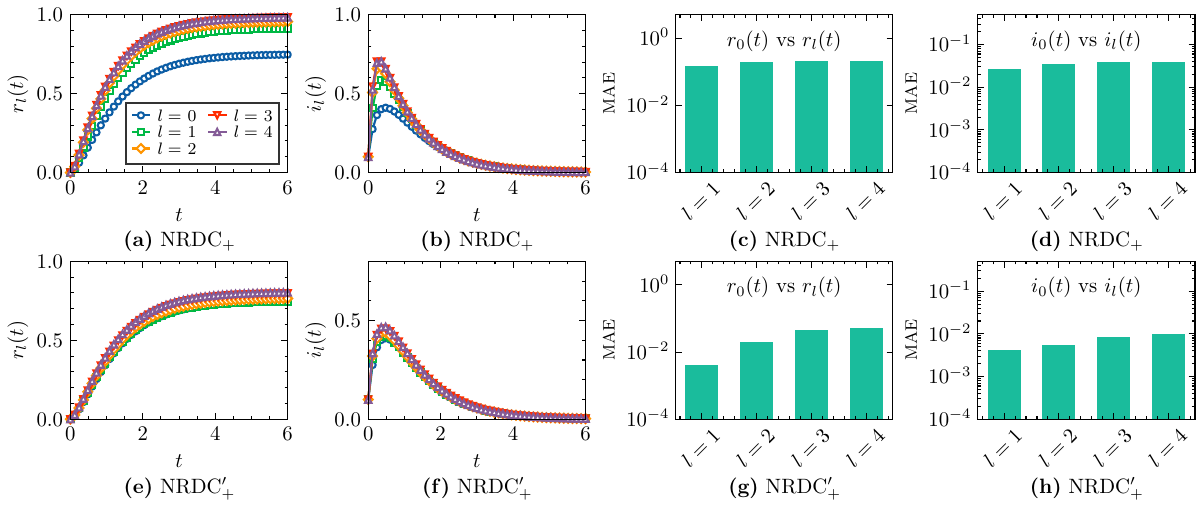}
\caption {The dependence of (a) the proportion of recovered nodes $r(t)$ and (b) the proportion of infected nodes $i(t)$ on the infection time $t$ in the DBLP network ($l=0$) and its four subnetworks ($l=1,2,3,4$), where the subnetworks are obtained via the NRDC$_{+}$ method. (c) and (d) show the mean absolute errors (MAE) between the epidemic dynamics curves of the corresponding subnetworks and the initial DBLP network. (e)-(h) show the results of the NRDC$_{+}^{\prime}$ reduction method.}
\label{Fig:S12}
\end{center}
\end{figure}

\begin{figure}[htbp]
\begin{center}
\includegraphics[width=1.0\linewidth]{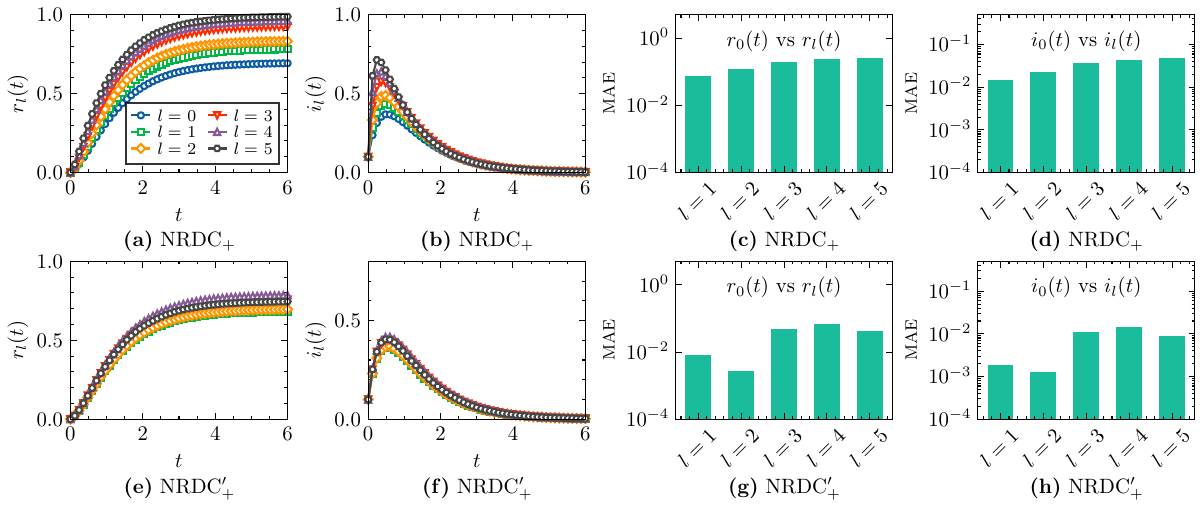}
\caption {The dependence of (a) the proportion of recovered nodes $r(t)$ and (b) the proportion of infected nodes $i(t)$ on the infection time $t$ in the Internet network ($l=0$) and its five subnetworks ($l=1,2,3,4,5$), where the subnetworks are obtained via the NRDC$_{+}$ method. (c) and (d) show the mean absolute errors (MAE) between the epidemic dynamics curves of the corresponding subnetworks and the initial Internet network. (e)-(h) show the results of the NRDC$_{+}^{\prime}$ reduction method.}
\label{Fig:S13}
\end{center}
\end{figure}

\begin{figure}[htbp]
\begin{center}
\includegraphics[width=1.0\linewidth]{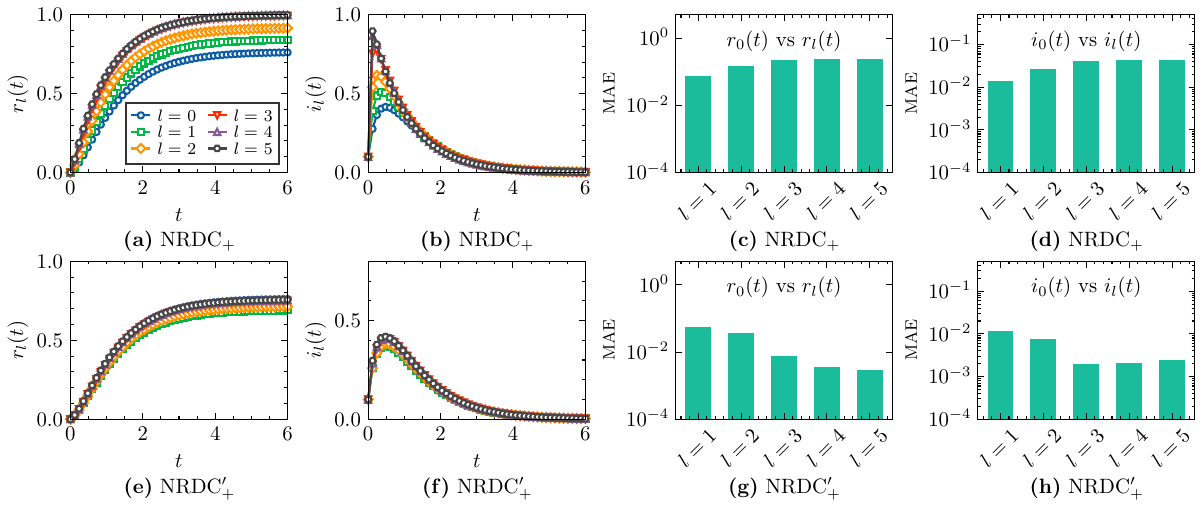}
\caption {The dependence of (a) the proportion of recovered nodes $r(t)$ and (b) the proportion of infected nodes $i(t)$ on the infection time $t$ in the Enron network ($l=0$) and its five subnetworks ($l=1,2,3,4,5$), where the subnetworks are obtained via the NRDC$_{+}$ method. (c) and (d) show the mean absolute errors (MAE) between the epidemic dynamics curves of the corresponding subnetworks and the initial Enron network. (e)-(h) show the results of the NRDC$_{+}^{\prime}$ reduction method.}
\label{Fig:S14}
\end{center}
\end{figure}

\begin{figure}[htbp]
\begin{center}
\includegraphics[width=0.6\linewidth]{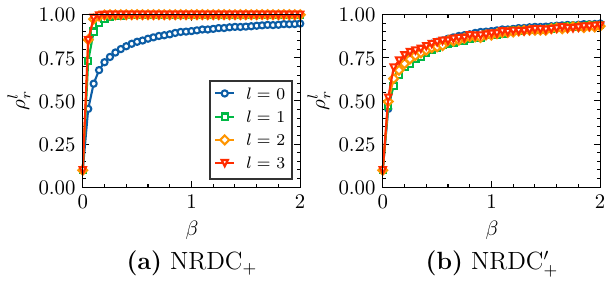}
\caption {The dependence of $\rho_r^l$ on the infection rate $\beta$ in the Blogs network ($l=0$) and its three subnetworks ($l=1,2,3$). (a) The subnetworks are obtained via the NRDC$_{+}$ method. (b) The subnetworks are obtained via the NRDC$_{+}^{\prime}$ method.}
\label{Fig:S15}
\end{center}
\end{figure}

\begin{figure}[htbp]
\begin{center}
\includegraphics[width=0.6\linewidth]{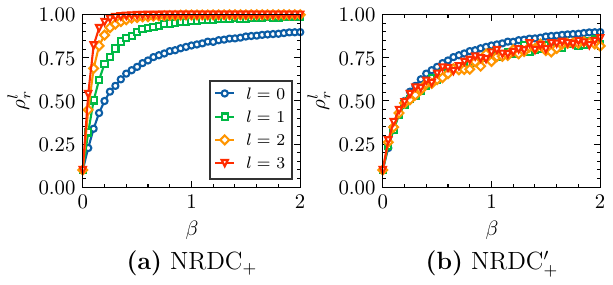}
\caption {The dependence of $\rho_r^l$ on the infection rate $\beta$ in the Drosophila network ($l=0$) and its three subnetworks ($l=1,2,3$). (a) The subnetworks are obtained via the NRDC$_{+}$ method. (b) The subnetworks are obtained via the NRDC$_{+}^{\prime}$ method.}
\label{Fig:S16}
\end{center}
\end{figure}

\begin{figure}[htbp]
\begin{center}
\includegraphics[width=0.6\linewidth]{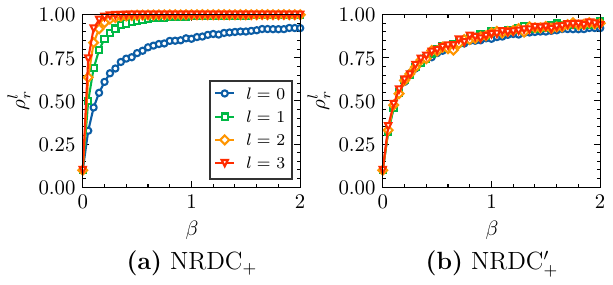}
\caption {The dependence of $\rho_r^l$ on the infection rate $\beta$ in the Music network ($l=0$) and its three subnetworks ($l=1,2,3$). (a) The subnetworks are obtained via the NRDC$_{+}$ method. (b) The subnetworks are obtained via the NRDC$_{+}^{\prime}$ method.}
\label{Fig:S17}
\end{center}
\end{figure}

\begin{figure}[htbp]
\begin{center}
\includegraphics[width=0.6\linewidth]{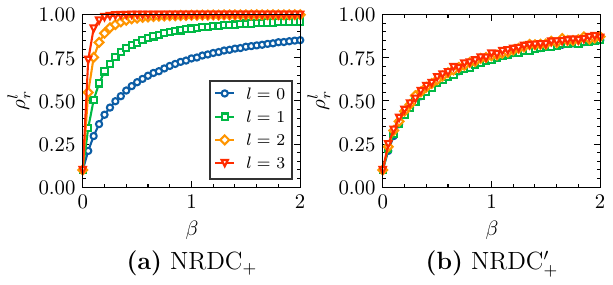}
\caption {The dependence of $\rho_r^l$ on the infection rate $\beta$ in the Airports network ($l=0$) and its three subnetworks ($l=1,2,3$). (a) The subnetworks are obtained via the NRDC$_{+}$ method. (b) The subnetworks are obtained via the NRDC$_{+}^{\prime}$ method.}
\label{Fig:S18}
\end{center}
\end{figure}

\begin{figure}[htbp]
\begin{center}
\includegraphics[width=0.6\linewidth]{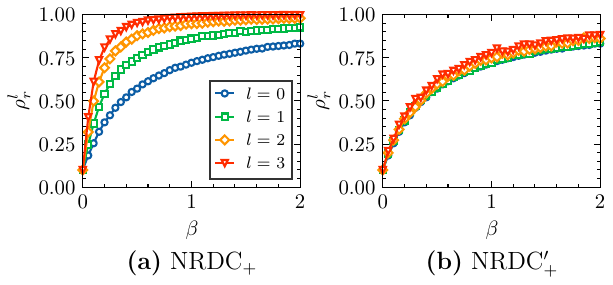}
\caption {The dependence of $\rho_r^l$ on the infection rate $\beta$ in the Proteome network ($l=0$) and its three subnetworks ($l=1,2,3$). (a) The subnetworks are obtained via the NRDC$_{+}$ method. (b) The subnetworks are obtained via the NRDC$_{+}^{\prime}$ method.}
\label{Fig:S19}
\end{center}
\end{figure}

\begin{figure}[htbp]
\begin{center}
\includegraphics[width=0.6\linewidth]{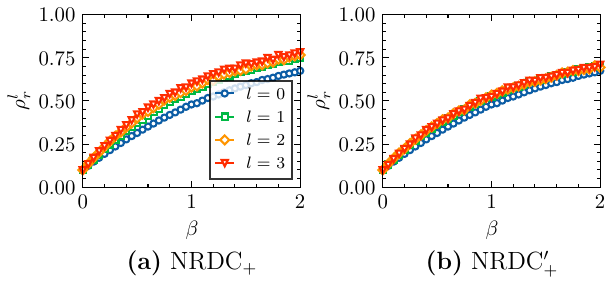}
\caption {The dependence of $\rho_r^l$ on the infection rate $\beta$ in the USpowergrid network ($l=0$) and its three subnetworks ($l=1,2,3$). (a) The subnetworks are obtained via the NRDC$_{+}$ method. (b) The subnetworks are obtained via the NRDC$_{+}^{\prime}$ method.}
\label{Fig:S20}
\end{center}
\end{figure}

\begin{figure}[htbp]
\begin{center}
\includegraphics[width=0.6\linewidth]{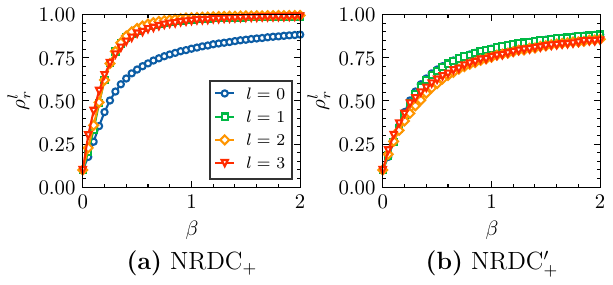}
\caption {The dependence of $\rho_r^l$ on the infection rate $\beta$ in the Gnutella network ($l=0$) and its three subnetworks ($l=1,2,3$). (a) The subnetworks are obtained via the NRDC$_{+}$ method. (b) The subnetworks are obtained via the NRDC$_{+}^{\prime}$ method.}
\label{Fig:S21}
\end{center}
\end{figure}

\begin{figure}[htbp]
\begin{center}
\includegraphics[width=0.6\linewidth]{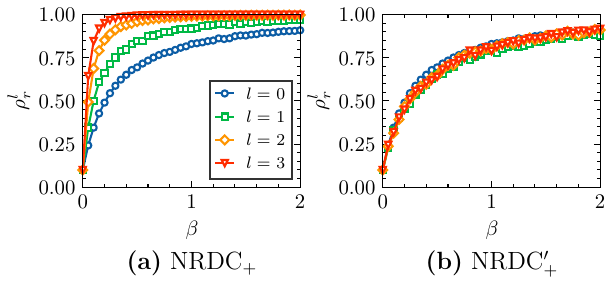}
\caption {The dependence of $\rho_r^l$ on the infection rate $\beta$ in the Words network ($l=0$) and its three subnetworks ($l=1,2,3$). (a) The subnetworks are obtained via the NRDC$_{+}$ method. (b) The subnetworks are obtained via the NRDC$_{+}^{\prime}$ method.}
\label{Fig:S22}
\end{center}
\end{figure}

\begin{figure}[htbp]
\begin{center}
\includegraphics[width=0.6\linewidth]{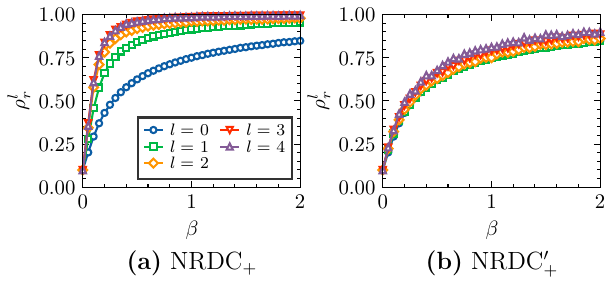}
\caption {The dependence of $\rho_r^l$ on the infection rate $\beta$ in the DBLP network ($l=0$) and its four subnetworks ($l=1,2,3,4$). (a) The subnetworks are obtained via the NRDC$_{+}$ method. (b) The subnetworks are obtained via the NRDC$_{+}^{\prime}$ method.}
\label{Fig:S23}
\end{center}
\end{figure}

\begin{figure}[htbp]
\begin{center}
\includegraphics[width=0.6\linewidth]{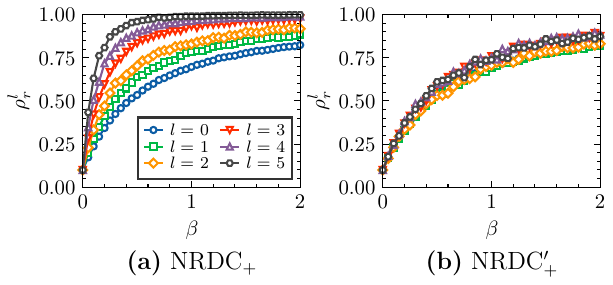}
\caption {The dependence of $\rho_r^l$ on the infection rate $\beta$ in the Internet network ($l=0$) and its five subnetworks ($l=1,2,3,4,5$). (a) The subnetworks are obtained via the NRDC$_{+}$ method. (b) The subnetworks are obtained via the NRDC$_{+}^{\prime}$ method.}
\label{Fig:S24}
\end{center}
\end{figure}

\begin{figure}[htbp]
\begin{center}
\includegraphics[width=0.6\linewidth]{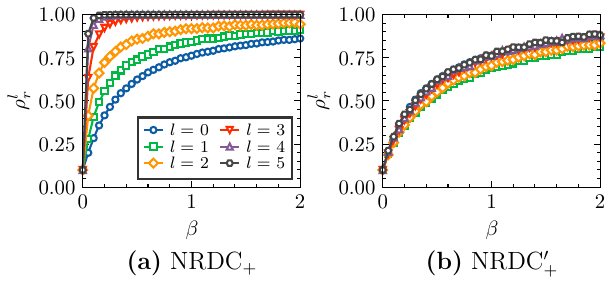}
\caption {The dependence of $\rho_r^l$ on the infection rate $\beta$ in the Enron network ($l=0$) and its five subnetworks ($l=1,2,3,4,5$). (a) The subnetworks are obtained via the NRDC$_{+}$ method. (b) The subnetworks are obtained via the NRDC$_{+}^{\prime}$ method.}
\label{Fig:S25}
\end{center}
\end{figure}

\newpage

\setcounter{figure}{0}
\renewcommand{\thefigure}{D\arabic{figure}}

\section{The information flow of real-world networks during the NRDC$_{+}$ and NRDC$_{+}^{\prime}$ processes}
We study the information flow in real-world networks (i.e., normalized partition function), as shown in Figures D1-D12. The results demonstrate that subnetworks obtained by the NRDC$_{+}^{\prime}$ method can better reproduce the information flow observed in original networks.

\begin{figure}[!h]
\begin{center}
\includegraphics[width=0.6\linewidth]{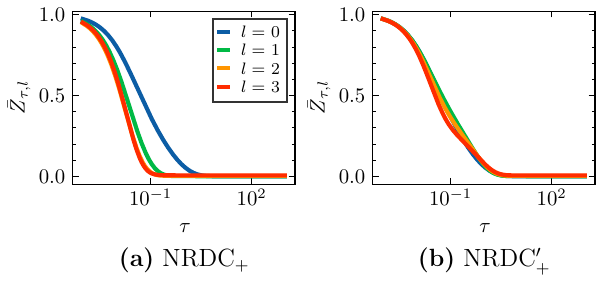}
\caption {The normalized partition functions of the Blogs network ($l=0$) and its three subnetworks ($l=1,2,3$). (a) The subnetworks are obtained via the NRDC$_{+}$ method. (b) The subnetworks are obtained via the NRDC$_{+}^{\prime}$ method.}
\label{Fig:S26}
\end{center}
\end{figure}

\begin{figure}[!h]
\begin{center}
\includegraphics[width=0.6\linewidth]{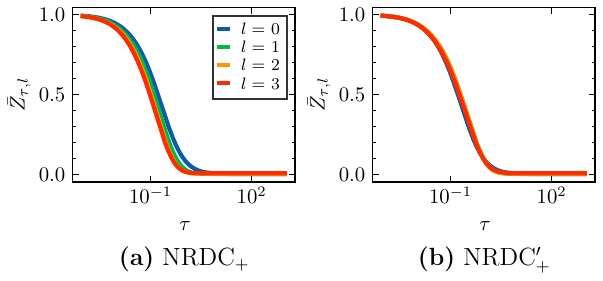}
\caption {The normalized partition functions of the Metabolic network ($l=0$) and its three subnetworks ($l=1,2,3$). (a) The subnetworks are obtained via the NRDC$_{+}$ method. (b) The subnetworks are obtained via the NRDC$_{+}^{\prime}$ method.}
\label{Fig:S27}
\end{center}
\end{figure}

\begin{figure}[!h]
\begin{center}
\includegraphics[width=0.6\linewidth]{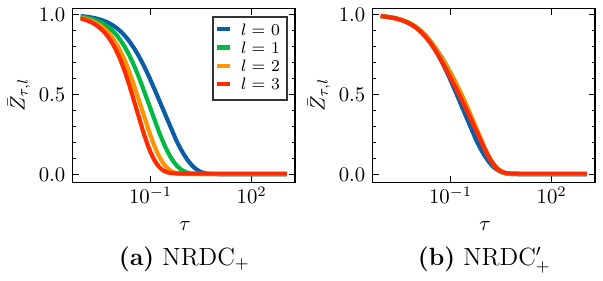}
\caption {The normalized partition functions of the Drosophila network ($l=0$) and its three subnetworks ($l=1,2,3$). (a) The subnetworks are obtained via the NRDC$_{+}$ method. (b) The subnetworks are obtained via the NRDC$_{+}^{\prime}$ method.}
\label{Fig:S28}
\end{center}
\end{figure}

\begin{figure}[!h]
\begin{center}
\includegraphics[width=0.6\linewidth]{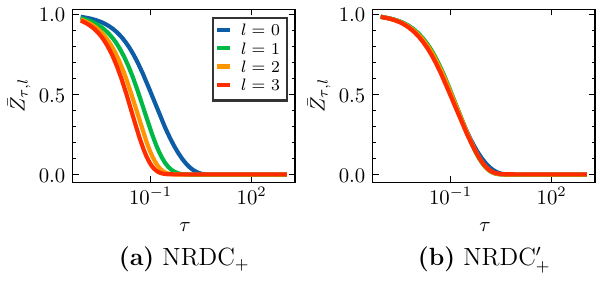}
\caption {The normalized partition functions of the Music network ($l=0$) and its three subnetworks ($l=1,2,3$). (a) The subnetworks are obtained via the NRDC$_{+}$ method. (b) The subnetworks are obtained via the NRDC$_{+}^{\prime}$ method.}
\label{Fig:S29}
\end{center}
\end{figure}

\begin{figure}[!h]
\begin{center}
\includegraphics[width=0.6\linewidth]{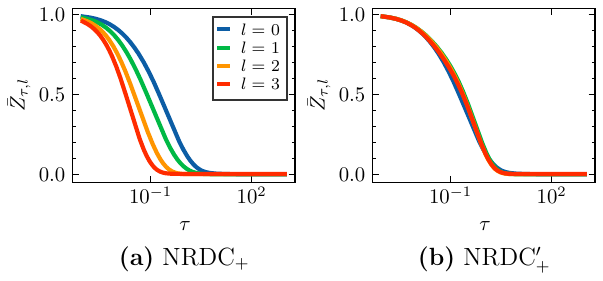}
\caption {The normalized partition functions of the Airports network ($l=0$) and its three subnetworks ($l=1,2,3$). (a) The subnetworks are obtained via the NRDC$_{+}$ method. (b) The subnetworks are obtained via the NRDC$_{+}^{\prime}$ method.}
\label{Fig:S30}
\end{center}
\end{figure}

\begin{figure}[!h]
\begin{center}
\includegraphics[width=0.6\linewidth]{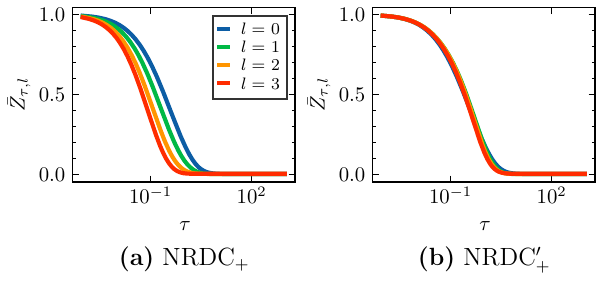}
\caption {The normalized partition functions of the Proteome network ($l=0$) and its three subnetworks ($l=1,2,3$). (a) The subnetworks are obtained via the NRDC$_{+}$ method. (b) The subnetworks are obtained via the NRDC$_{+}^{\prime}$ method.}
\label{Fig:S31}
\end{center}
\end{figure}

\begin{figure}[!h]
\begin{center}
\includegraphics[width=0.6\linewidth]{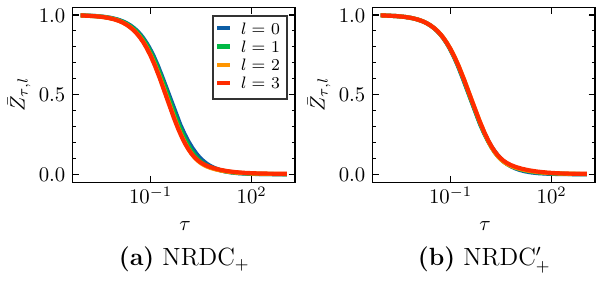}
\caption {The normalized partition functions of the USpowergrid network ($l=0$) and its three subnetworks ($l=1,2,3$). (a) The subnetworks are obtained via the NRDC$_{+}$ method. (b) The subnetworks are obtained via the NRDC$_{+}^{\prime}$ method.}
\label{Fig:S32}
\end{center}
\end{figure}

\begin{figure}[!h]
\begin{center}
\includegraphics[width=0.6\linewidth]{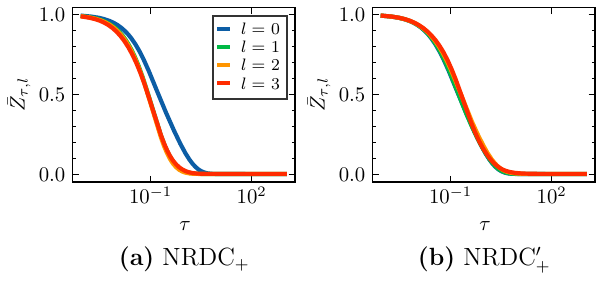}
\caption {The normalized partition functions of the Gnutella network ($l=0$) and its three subnetworks ($l=1,2,3$). (a) The subnetworks are obtained via the NRDC$_{+}$ method. (b) The subnetworks are obtained via the NRDC$_{+}^{\prime}$ method.}
\label{Fig:S33}
\end{center}
\end{figure}

\begin{figure}[!h]
\begin{center}
\includegraphics[width=0.6\linewidth]{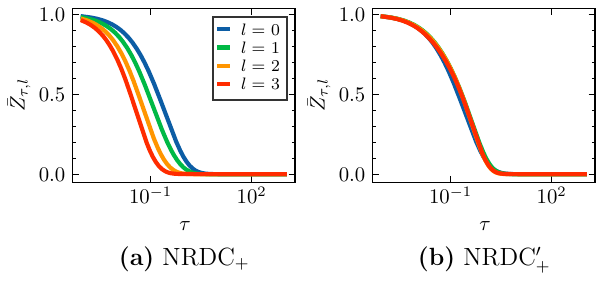}
\caption {The normalized partition functions of the Words network ($l=0$) and its three subnetworks ($l=1,2,3$). (a) The subnetworks are obtained via the NRDC$_{+}$ method. (b) The subnetworks are obtained via the NRDC$_{+}^{\prime}$ method.}
\label{Fig:S34}
\end{center}
\end{figure}

\begin{figure}[!h]
\begin{center}
\includegraphics[width=0.6\linewidth]{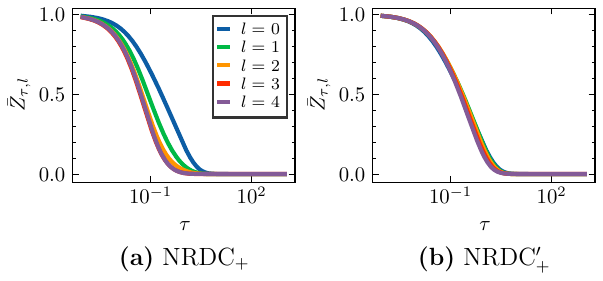}
\caption {The normalized partition functions of the DBLP network ($l=0$) and its four subnetworks ($l=1,2,3,4$). (a) The subnetworks are obtained via the NRDC$_{+}$ method. (b) The subnetworks are obtained via the NRDC$_{+}^{\prime}$ method.}
\label{Fig:S35}
\end{center}
\end{figure}

\begin{figure}[!h]
\begin{center}
\includegraphics[width=0.6\linewidth]{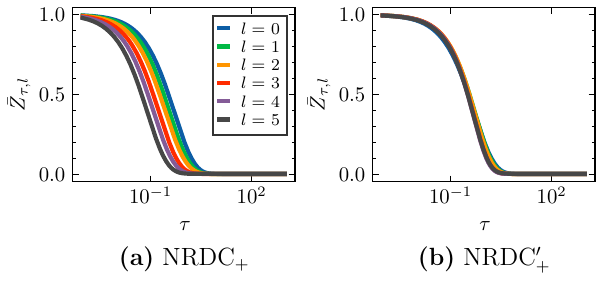}
\caption {The normalized partition functions of the Internet network ($l=0$) and its five subnetworks ($l=1,2,3,4,5$). (a) The subnetworks are obtained via the NRDC$_{+}$ method. (b) The subnetworks are obtained via the NRDC$_{+}^{\prime}$ method.}
\label{Fig:S36}
\end{center}
\end{figure}

\begin{figure}[!h]
\begin{center}
\includegraphics[width=0.6\linewidth]{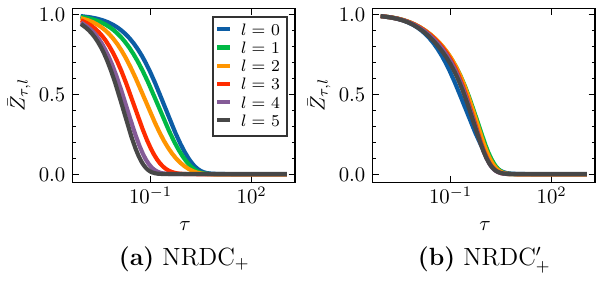}
\caption {The normalized partition functions of the Enron network ($l=0$) and its five subnetworks ($l=1,2,3,4,5$). (a) The subnetworks are obtained via the NRDC$_{+}$ method. (b) The subnetworks are obtained via the NRDC$_{+}^{\prime}$ method.}
\label{Fig:S37}
\end{center}
\end{figure}

\clearpage

\begin{figure*}[!h]
\begin{center}
\includegraphics[width=0.6\linewidth]{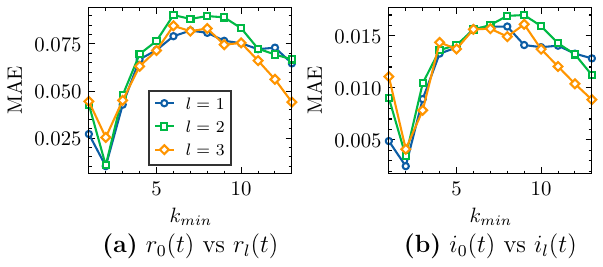}
\caption {The dependence of the mean absolute errors (MAE) between the epidemic dynamics curves of the Music network and its three subnetworks ($l=1,2,3$) on $k_{min}$, where the subnetworks are obtained via the NRDC$_{+}^{\prime}$ method.}
\label{Fig:S38}
\end{center}
\end{figure*}

\newpage

\setcounter{figure}{0}
\renewcommand{\thefigure}{E\arabic{figure}}

\setcounter{table}{0}
\renewcommand{\thetable}{E\arabic{table}}

\section{Comparison of the NRDC$_{+}^{\prime}$ method with other mainstream reduction methods}
To further demonstrate the superiority of the NRDC$_{+}^{\prime}$ method, we conducted a series of comparative analyses against several mainstream network reduction techniques, including four subgraph sampling methods and three renormalization methods. Figures E1 and E2 present the results comparing the NRDC$_{+}$ method with the four sampling methods on BA scale-free networks. In addition, similar to Table 2 in the main text, Tables E1 and E2 provide the $f_{overlap}$ values of real-world networks under the other two reduction rates. Figures E3-E10 show the comparison results between our method and the three renormalization methods. The results indicate that NRDC$_{+}^{\prime}$ not only better preserves the spreading dynamics and information flow characteristics of the original network but also has lower computational complexity, as shown in Figure E11. It should be noted that in pruning Algorithm 1, NRDC$_{+}^{\prime}$ significantly outperforms the renormalization methods, whether $k_{min}$ is set to the optimal value based on the guidance from Figure D12 or simply set to 2. For each method, the number of nodes in the reduced network is 1/8 of that of the original network.

\begin{figure*}[!h]
\centering
\includegraphics[width=0.6\linewidth]{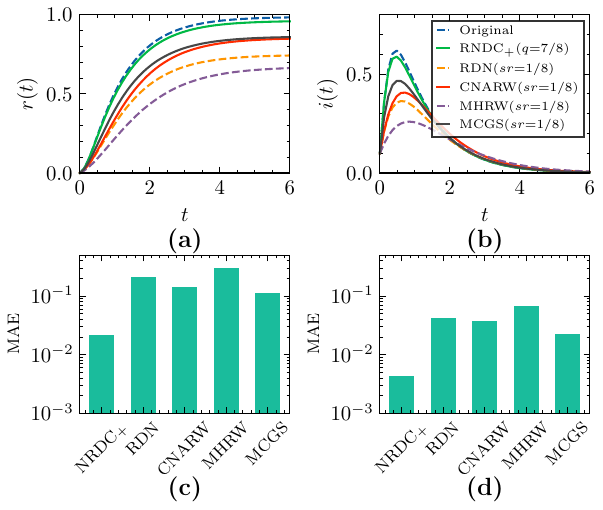}
\caption{Comparison of epidemic dynamics behaviors of the BA network under five reduction methods. (a) and (b) show the epidemic spreading dynamics curves of the original BA network and the target subgraphs, where the subgraphs are obtained via the five methods of NRDC$_{+}$ (the node removal ratio is $q = 7/8$), RDN (the node sampling rate is $sr = 1/8$), CNARW ($sr = 1/8$), MHRW ($sr = 1/8$), and MCGS ($sr = 1/8$) methods, respectively. In each case, the node size of the target subgraph is 1/8 of that of the original network. (c) and (d) show the mean absolute errors (MAE) between the epidemic dynamics curves of the target subgraphs and the original network under the corresponding methods. The methods RDN, CNARW, and MHRW can be implemented via the Python third-party library Little Ball of Fur: \url{https://github.com/benedekrozemberczki/littleballoffur}. The open-source code for MCGS is available at \url{https://github.com/csuvis/MCGS}.}
\label{Fig:S39}
\end{figure*}

\begin{figure*}[!h]
\centering
\includegraphics[width=0.6\linewidth]{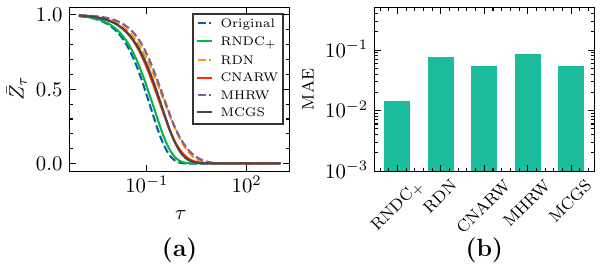}
\caption{Comparison of information flows of the BA scale-free networks under five reduction methods. (a) The partition function curves of the original BA network and the target subgraphs, where the subgraphs are obtained via the five methods of NRDC$_{+}$ (the node removal ratio is $q = 7/8$), RDN (the node sampling rate is $sr = 1/8$), CNARW ($sr = 1/8$), MHRW ($sr = 1/8$), and MCGS ($sr = 1/8$) methods, respectively. In each case, the node size of the target subgraph is 1/8 of that of the original network. (b) The mean absolute errors (MAE) between the partition function curves of the target subgraphs and the original network.}
\label{Fig:S40}
\end{figure*}

%============================== Table S1========================
\begin{table}[!h]
\footnotesize
\centering
\caption{The value of $f_{overlap}$ is determined by comparing the spreading ability curves $\rho_r^0(\beta)$ of the initial network $G_0$ with $\rho_r^l(\beta)$ of the corresponding subnetwork $G_l$ ($l=1$). For four baseline sampling methods (RDN, CNARW, MHRW, and MCGS), the node sampling rate $sr = 1/2$. All the best results are highlighted in bold, and the second-best results are marked with an underline.}
\label{table:S1}
\tabcolsep 12pt %space between two columns. 用于调整列间距
\begin{tabular*}{\textwidth}{ccccccccc}
\toprule
Name & $l$ & $k_{min}$ & RDN & CNARW & MHRW & MCGS &NRDC$_{+}$ &NRDC$_{+}^{\prime}$  \\
\hline
ER($N=5000$) &1 &***   &0.8374     &0.8819  &0.8721  &\underline{0.8899} &$\mathbf{0.9635}$  &*** \\
BA($N=5000$) &1 &***   &0.9097  &0.9403   &0.9014  &\underline{0.9514} &$\mathbf{0.9642}$   &*** \\
Blogs   &1 &2   &0.8424  &0.8447   &0.8492   &\underline{0.8913}    &0.8079   &$\mathbf{0.9602}$ \\
Metabolic   &1 &2   &\underline{0.9502} &0.9227 &0.9255   &0.9104   &0.8408   &$\mathbf{0.9782}$ \\
Drosophila  &1 &2   &0.8384 &\underline{0.8509} &0.8436   &0.8282   &0.7784            &$\mathbf{0.9008}$ \\
Music       &1 &2   &0.8263 &\underline{0.8941} &0.8110   &0.8696   &0.7851            &$\mathbf{0.9606}$ \\
Airports    &1 &2   &0.8134 &\underline{0.8307} &0.8181   &0.7743   &0.7422            &$\mathbf{0.9919}$ \\
Proteome    &1 &2   &0.8220 &0.8305 &\underline{0.8392}   &0.8329   &0.7897            &$\mathbf{0.9871}$ \\
USpowergrid &1 &2 &0.9416 &$\mathbf{0.9929}$ &\underline{0.9855}   &0.9509   &0.8834 &0.9263 \\
Gnutella &1 &2 &0.8534  &0.8538   &\underline{0.8712}  &0.8581    &0.7608  &$\mathbf{0.9931}$ \\
Words       &1 &2   &0.8580 &0.8674 &0.8496   &\underline{0.8931}   &0.8269         &$\mathbf{0.9336}$ \\
DBLP &1 &2 &0.7939  &0.8043        &\underline{0.8063}  &0.7954    &0.7554    &$\mathbf{0.9931}$ \\
Internet    &1 &2   &0.8823 &\underline{0.9242} &0.8967   &0.8763   &0.8717          &$\mathbf{0.9856}$ \\
Enron       &1 &2   &0.8465 &0.8829 &0.7925   &\underline{0.8833}   &0.8538         &$\mathbf{0.8841}$ \\
\bottomrule
\end{tabular*}
\end{table}
%==========================  end Table S1 ======================

%============================== Table S2========================
\begin{table}[!h]
\footnotesize
\centering
\caption{The value of $f_{overlap}$ is determined by comparing the spreading ability curves $\rho_r^0(\beta)$ of the initial network $G_0$ with $\rho_r^l(\beta)$ of the corresponding subnetwork $G_l$ ($l=2$). For four baseline sampling methods (RDN, CNARW, MHRW, and MCGS), the node sampling rate $sr = 1/4$. All the best results are highlighted in bold, and the second-best results are marked with an underline.}
\label{table:S2}
\tabcolsep 12pt %space between two columns. 用于调整列间距
\begin{tabular*}{\textwidth}{ccccccccc}
\toprule
Name        &$l$ &$k_{min}$ &RDN &CNARW &MHRW &MCGS &NRDC$_{+}$ &NRDC$_{+}^{\prime}$  \\
\hline
ER($N=5000$) &2 &***   &0.6032             &0.7280              &0.7151           &\underline{0.7396          } &$\mathbf{0.8564}$            &*** \\
BA($N=5000$) &2 &***   &0.7865             &0.8572              &0.7574           &\underline{0.8724          } &$\mathbf{0.9430}$            &*** \\
Blogs   &2 &2   &0.8387  &0.8454   &0.8557   &\underline{0.8901}    &0.7985   &$\mathbf{0.9641}$ \\
Metabolic   &2   &2   &$\mathbf{0.9819}$    &0.9321    &\underline{0.9724}    &0.9293    &0.7900      &0.9677 \\
Drosophila  &2   &2   &0.8047    &0.8022    &\underline{0.8704}    &0.8552         &0.7143     &$\mathbf{0.8895}$ \\
Music       &2   &2   &0.8033    &\underline{0.8560}    &0.7852    &0.8511        &0.7544      &$\mathbf{0.9653}$ \\
Airports    &2   &2   &0.7378    &0.7449    &\underline{0.8037}    &0.7193            &0.6467  &$\mathbf{0.9624}$ \\
Proteome    &2   &2   &0.7866    &0.7954    &0.8086    &\underline{0.8283}            &0.6902  &$\mathbf{0.9534}$ \\
USpowergrid &2   &2 &0.8460    &$\mathbf{0.9936}$    &\underline{0.9808}    &0.7435     &0.8564 &0.9212 \\
Gnutella &2 &2 &$\mathbf{0.9174}$  &0.8971   &0.9058  &0.8858    &0.7504  &\underline{0.9136} \\
Words       &2   &2   &0.8096    &0.8220    &0.8080    &\underline{0.8635}            &0.7288  &$\mathbf{0.9638}$ \\
DBLP &2 &2 &0.7541  &0.7566        &\underline{0.7823}  &0.7709    &0.6975    &$\mathbf{0.9755}$ \\
Internet    &2   &2   &0.8316    &\underline{0.8872}    &0.8535    &0.8383            &0.7941  &$\mathbf{0.9779}$ \\
Enron       &2   &2   &0.8001    &0.7958    &0.7584    &\underline{0.8314}            &0.7462  &$\mathbf{0.9119}$ \\
\bottomrule
\end{tabular*}
\end{table}
%==========================  end Table S2======================

\begin{figure*}[!h]
\centering
\includegraphics[width=0.6\linewidth]{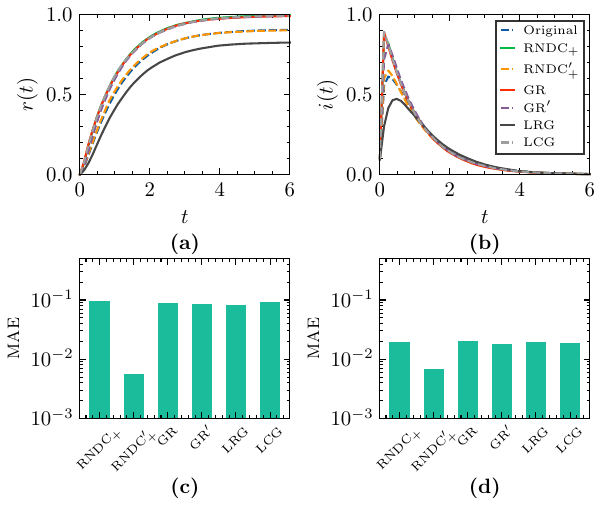}
\caption{Comparison of epidemic dynamics in the Blogs network under the NRDC$_{+}^{\prime}$ (in Algorithm 1, the parameter $k_{min}=3$) and several renormalization methods. (a) and (b) display the epidemic dynamics curves of the original Blogs network and the reduced networks. (c) and (d) show the mean absolute errors (MAE) between the epidemic dynamics curves of the reduced networks and those of the original Blogs network.}
\label{Fig:S41}
\end{figure*}

\begin{figure*}[!h]
\centering
\includegraphics[width=0.6\linewidth]{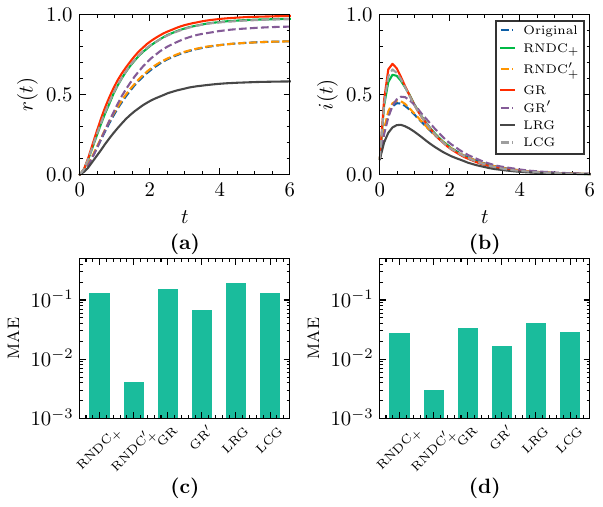}
\caption{Comparison of epidemic dynamics in the Metabolic network under the NRDC$_{+}^{\prime}$ (in Algorithm 1, the parameter $k_{min}=13$) and several renormalization methods. (a) and (b) display the epidemic dynamics curves of the original Metabolic network and the reduced networks. (c) and (d) show the mean absolute errors (MAE) between the epidemic dynamics curves of the reduced networks and those of the original Metabolic network.}
\label{Fig:S42}
\end{figure*}

\begin{figure*}[!h]
\centering
\includegraphics[width=0.6\linewidth]{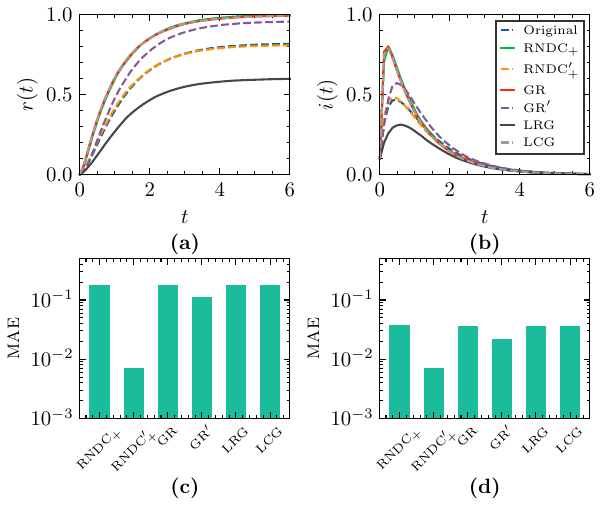}
\caption{Comparison of epidemic dynamics in the Drosophila network under the NRDC$_{+}^{\prime}$ (in Algorithm 1, the parameter $k_{min}=13$) and several renormalization methods. (a) and (b) display the epidemic dynamics curves of the original Drosophila network and the reduced networks. (c) and (d) show the mean absolute errors (MAE) between the epidemic dynamics curves of the reduced networks and those of the original Drosophila network.}
\label{Fig:S43}
\end{figure*}

\begin{figure*}[!h]
\centering
\includegraphics[width=0.6\linewidth]{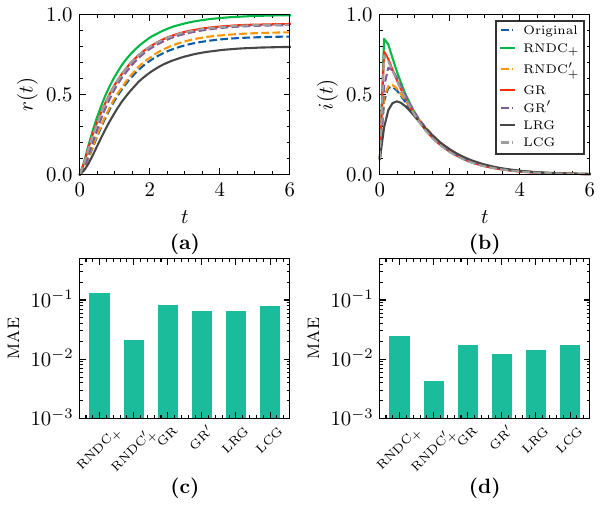}
\caption{Comparison of epidemic dynamics in the Music network under the NRDC$_{+}^{\prime}$ (in Algorithm 1, the parameter $k_{min}=2$) and several renormalization methods. (a) and (b) display the epidemic dynamics curves of the original Music network and the reduced networks. (c) and (d) show the mean absolute errors (MAE) between the epidemic dynamics curves of the reduced networks and those of the original Music network.}
\label{Fig:S44}
\end{figure*}

\begin{figure*}[!h]
\centering
\includegraphics[width=0.7\linewidth]{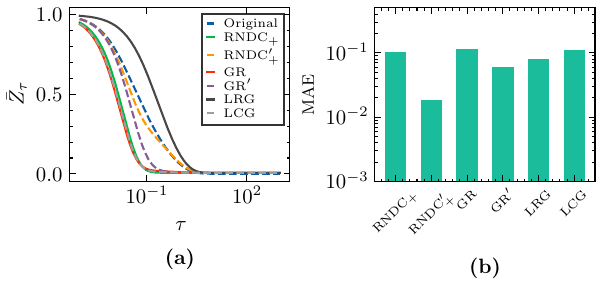}
\caption{Comparison of normalized partition functions in the Blogs network under the NRDC$_{+}^{\prime}$ (in Algorithm 1, the parameter $k_{min}=3$) and several renormalization methods. (a) and (b) display the normalized partition functions curves of the original Blogs network and the reduced networks. (c) and (d) show the mean absolute errors (MAE) between the normalized partition functions curves of the reduced networks and those of the original Blogs network.}
\label{Fig:S45}
\end{figure*}

\begin{figure*}[!h]
\centering
\includegraphics[width=0.7\linewidth]{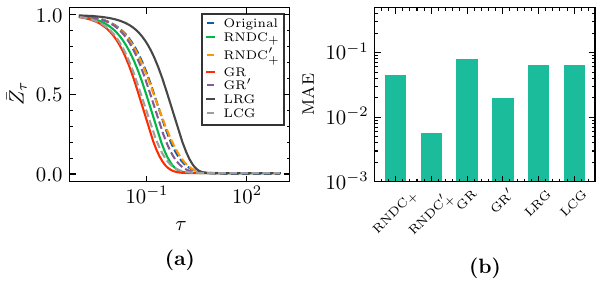}
\caption{Comparison of normalized partition functions in the Metabolic network under the NRDC$_{+}^{\prime}$ (in Algorithm 1, the parameter $k_{min}=13$) and several renormalization methods. (a) and (b) display the normalized partition functions curves of the original Metabolic network and the reduced networks. (c) and (d) show the mean absolute errors (MAE) between the normalized partition functions curves of the reduced networks and those of the original Metabolic network.}
\label{Fig:S46}
\end{figure*}

\begin{figure*}[!h]
\centering
\includegraphics[width=0.7\linewidth]{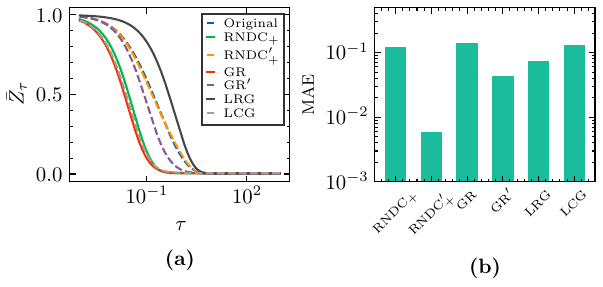}
\caption{Comparison of normalized partition functions in the Drosophila network under the NRDC$_{+}^{\prime}$ (in Algorithm 1, the parameter $k_{min}=13$) and several renormalization methods. (a) and (b) display the normalized partition functions curves of the original Drosophila network and the reduced networks. (c) and (d) show the mean absolute errors (MAE) between the normalized partition functions curves of the reduced networks and those of the original Drosophila network.}
\label{Fig:S47}
\end{figure*}

\begin{figure*}[!h]
\centering
\includegraphics[width=0.7\linewidth]{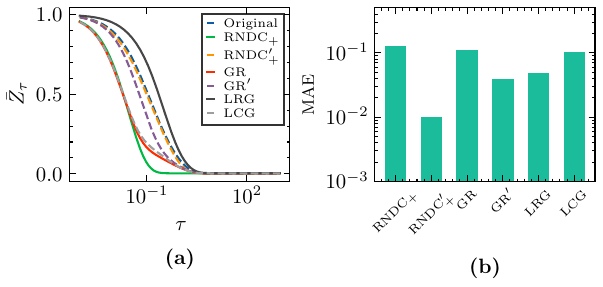}
\caption{Comparison of normalized partition functions in the Music network under the NRDC$_{+}^{\prime}$ (in Algorithm 1, the parameter $k_{min}=2$) and several renormalization methods. (a) and (b) display the normalized partition functions curves of the original Music network and the reduced networks. (c) and (d) show the mean absolute errors (MAE) between the normalized partition functions curves of the reduced networks and those of the original Music network.}
\label{Fig:S48}
\end{figure*}

\begin{figure*}[!h]
\centering
\includegraphics[width=1.0\linewidth]{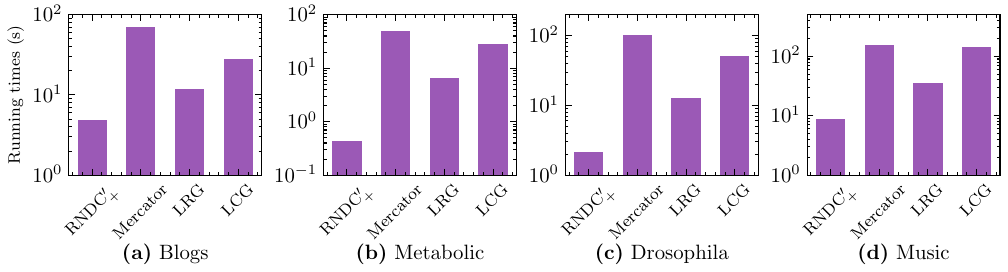}
\caption{The running times for reducing the number of nodes in the four real-world networks to half of the original using the NRDC$_{+}^{\prime}$ and three renormalization methods (GR, LRG, and LCG), respectively. Here, Mercator (\url{https://github.com/networkgeometry/mercator}) serves as the core step of geometric renormalization (GR), which is used to embed the topological structure of the network into hyperbolic space.}
\label{Fig:S49}
\end{figure*}

\end{appendix}

\end{document}